\documentclass[11pt]{article}
\usepackage{amssymb}
\usepackage[final]{acl}

\usepackage{times}
\usepackage{latexsym}
\usepackage{pifont}
\newcommand{\cmark}{\ding{51}}   % ✓
\newcommand{\xmark}{\ding{55}}   % ✗
\usepackage[table]{xcolor}   % 必须加 [table]
 \usepackage{amsmath}
\definecolor{lightgray}{gray}{0.92}
\usepackage{booktabs}
\usepackage{ulem} % for \uline
\usepackage[most]{tcolorbox}
\usepackage{enumitem} % 用于更好地控制列表间距

\definecolor{boxTitleBg}{RGB}{80, 120, 160}    % 标题栏深蓝色背景
\definecolor{boxContentBg}{RGB}{240, 245, 250} % 内容区域浅蓝灰色背景
\definecolor{boxFrameColor}{RGB}{80, 120, 160} % 边框颜色（与标题栏一致）

\newtcolorbox{systemprompt}[1][]{
  enhanced,               % 启用高级功能
  title={System Instruction}, % 固定标题
  colframe=boxFrameColor, % 边框颜色
  colbacktitle=boxTitleBg,% 标题栏背景色
  coltitle=white,         % 标题文字颜色
  colback=boxContentBg,   % 内容背景色
  fonttitle=\bfseries\rmfamily, % 标题字体：加粗 + 衬线体(Times)
  fontupper=\rmfamily,    % 【关键】内容字体：强制使用衬线体(Times)，与正文一致
  arc=1.5mm,              % 圆角半径
  boxrule=0.7pt,          % 边框线条粗细
  titlerule=0pt,          % 标题和内容之间不需要分割线，靠颜色区分
  left=5pt, right=5pt, top=5pt, bottom=5pt, % 内部边距
  #1 % 可选参数用于微调
}

\usepackage{makecell}
\usepackage{array}
\usepackage{multirow} % 必须引入，用于合并行实现垂直居中

\usepackage[T1]{fontenc}
\usepackage{natbib}

\usepackage[utf8]{inputenc}

\usepackage{microtype}

\usepackage{inconsolata}

\usepackage{graphicx}

\usepackage{xcolor} % 必须引入颜色宏包

\title{MTAVG-Bench: A Diagnostic Benchmark for Multi-Talker Dialogue-Centric Audio-Video Generation}

\author{
 \textbf{Yang-Hao Zhou\textsuperscript{1,†,‡}},
 \textbf{Haitian Li\textsuperscript{2,†}},
 \textbf{Rexar Lin\textsuperscript{1,†}},
\\
 \textbf{Heyan Huang\textsuperscript{1,*}},
 \textbf{Jinxing Zhou\textsuperscript{3}},
 \textbf{Changsen Yuan\textsuperscript{4}},
 \textbf{Tian Lan\textsuperscript{1}},
 \textbf{Ziqin Zhou\textsuperscript{5}},
\\
 \textbf{Yudong Li\textsuperscript{6}},
 \textbf{Jiajun Xu\textsuperscript{7}},
 \textbf{Jingyun Liao\textsuperscript{7}},
 \textbf{Yiming Cheng\textsuperscript{6}},
\\
 \textbf{Xuefeng Chen\textsuperscript{7}},
 \textbf{Xian-Ling Mao\textsuperscript{1}},
 \textbf{Yousheng Feng\textsuperscript{7}}
\\
 \textsuperscript{1}Beijing Institute of Technology,
 \textsuperscript{2}Shanghai University,
 \textsuperscript{3}OpenNLP Lab,
 \textsuperscript{4}Beijing University of Technology,
\\
 \textsuperscript{5}The University of Adelaide,
 \textsuperscript{6}Tsinghua University,
 \textsuperscript{7}Inkeverse Group Limited.
\\
\textsuperscript{*}Corresponding author.
\textsuperscript{†}Equal Contribution.
\textsuperscript{‡}Project Leader.
}

\begin{document}
\maketitle
\begin{abstract}
Recent advances in text-to-audio-video (T2AV) generation have enabled models to synthesize audio-visual videos with multi-participant dialogues. However, existing evaluation benchmarks remain largely designed for human-recorded videos or single-speaker settings. As a result, structural failures in generated multi-talker dialogue videos, such as identity drift, unnatural turn transitions, and audio-visual misalignment, cannot be effectively diagnosed. To address this issue, we introduce \textbf{MTAVG-Bench}\footnote{Our benchmark is available at \url{https://mortyzhou77.github.io/MTAVG-Bench-project-page}}, a failure-driven diagnostic benchmark for multi-talker dialogue-centric audio-video generation. MTAVG-Bench is built via a semi-automatic pipeline, where 1.8k videos are generated using mainstream T2AV models with carefully designed prompts, yielding 2.4k manually annotated QA pairs for fine-grained failure diagnosis. The benchmark evaluates multi-speaker dialogue generation at four levels: audio-visual signal fidelity, temporal attribute consistency, social interaction, and cinematic expression. Built on a hierarchical failure taxonomy and a targeted QA protocol, MTAVG-Bench is primarily designed to evaluate whether proprietary and open-source omni-models can reliably identify failure modes in multi-speaker T2AV outputs. We benchmark 12 proprietary and open-source omni-models on MTAVG-Bench, with Gemini 3 Pro achieving the strongest overall performance, while leading open-source models remain competitive in signal fidelity and consistency. Overall, MTAVG-Bench enables fine-grained failure analysis for rigorous model comparison and targeted video generation refinement.

\end{abstract}

\begin{figure*}[t]
  \centering
  \includegraphics[width=\textwidth]{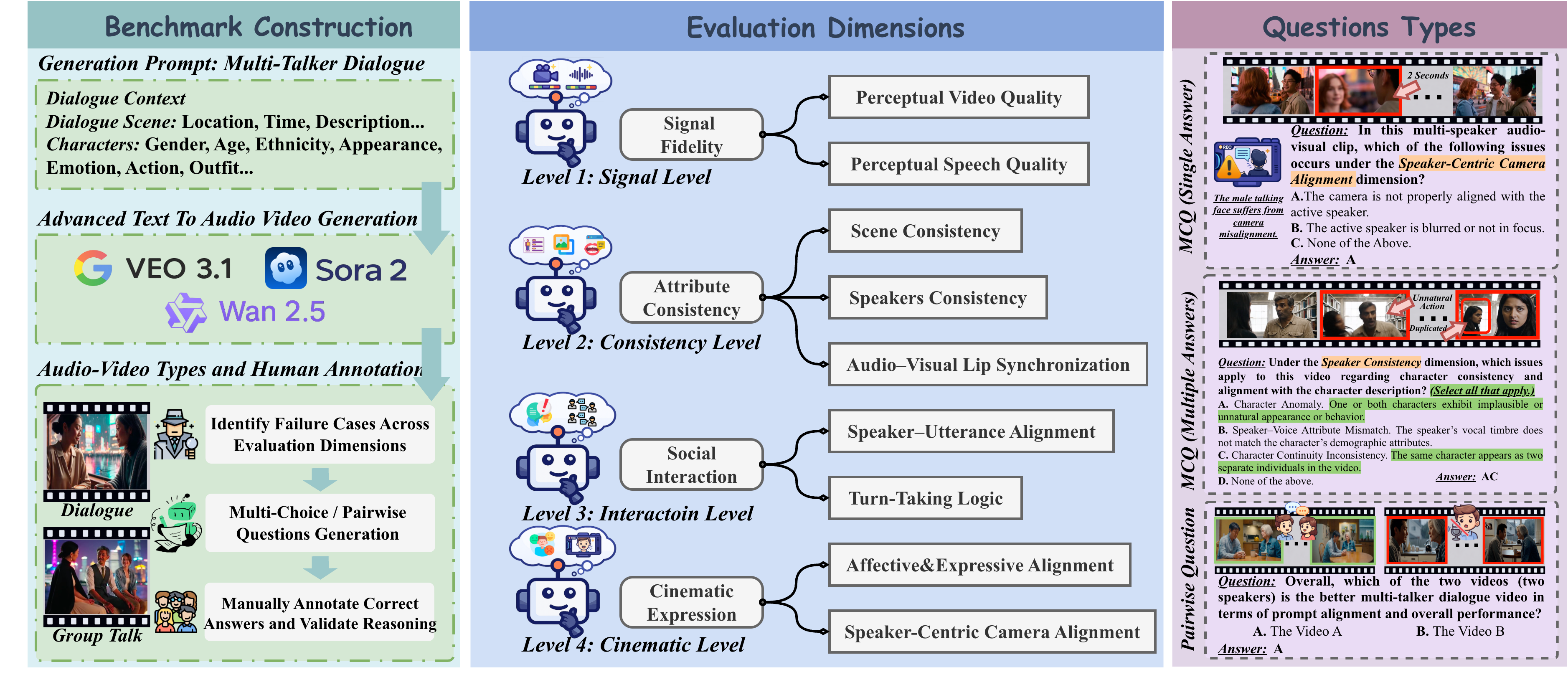}
\vspace{-5ex}
  \caption{MTAVG-Bench is a benchmark for evaluating text-to-audio-video (T2AV) models on multi-talker dialogue generation, built by synthesizing dialogue-driven videos from structured prompts and collecting human annotations based on carefully defined fine-grained evaluation dimensions. It features a four-level evaluation framework and diverse multi-choice and pairwise questions that assess signal quality, consistency, social interaction, and cinematic expression, with a focus on failure mode in cinematic speaker-centric dialogue video generation.}
  \label{fig:multitalkbench-overview}
\vspace{-2mm}
\end{figure*}

\section{Introduction}

Recent advances in text-to-audio-video (T2AV) generation have evolved from synthesizing simple environmental sounds to unified audio-visual content featuring natural speech~\cite{mao2024tavgbench, Javisdit,Uniavgen,low2025ovi,hacohen2026ltx,wang20251,team2026mova}. The emergence of high-fidelity commercial systems, such as Veo 3~\cite{google2025veo3}, Sora 2~\cite{openai2025sora2}, and Wan 2.5~\cite{alibaba2025wan25}, marks a significant transition toward movie-level production. To assess these growing capabilities, the research community has established benchmarks that primarily focus on general audio-visual events~\cite{Javisdit,JavisGPT,hua2025vabench} or specific single-speaker attributes like lip synchronization~\cite{zhang2024comparative,zhou2024thqa,nocentini2024beyond} and interactive talking face~\cite{zhou2022responsive,zhou2025interactive,zhu2025multi}.

However, these existing frameworks fail to address the structural complexities of multi-talker dialogue scenarios. In this domain, the primary challenge shifts from low-level perceptual fidelity to high-level structural coherence, such as maintaining speaker identity and logical turn-taking. Even state-of-the-art systems often produce visually realistic videos that nonetheless suffer from critical failures like identity drift and audio-visual misalignment. Since these errors stem from semantic reasoning and cross-modal consistency rather than perceptual quality, they are poorly captured by existing realism-oriented evaluation metrics.

To address the limitations of existing evaluations for multi-talker dialogue generation, we introduce \textbf{M}ulti-\textbf{T}alker \textbf{A}udio-\textbf{V}isual \textbf{G}eneration Benchmark (\textbf{MTAVG-Bench}), a diagnostic benchmark for multi-talker dialogue-centric audio-video generation. MTAVG-Bench aims to systematically characterize structural failures and cross-modal inconsistencies in multi-speaker dialogue scenarios, while providing a foundation for downstream video diagnosis, editing, and refinement. As illustrated in Figure~\ref{fig:multitalkbench-overview}, the benchmark first expands text-based dialogue prompts into dialogue-centric generation prompts that specify dialogue content, scene context, and speaker attributes, which are then used to synthesize multi-talker dialogue videos with a diverse set of state-of-the-art T2AV systems. The generated videos are subsequently annotated with fine-grained human labels that capture typical generation errors and distortions in multi-talker audio-visual dialogue.

MTAVG-Bench further adopts a four-level evaluation framework covering \textit{Signal Fidelity}, \textit{Attribute Consistency}, \textit{Social Interaction}, and \textit{Cinematic Expression}, which progressively characterize multi-talker dialogue generation from low-level perceptual quality to high-level structural and cinematic coherence. 
Each level is decomposed into fine-grained, diagnosable dimensions, including perceptual video quality, perceptual speech quality, scene consistency, speaker consistency, audio-visual lip synchronization, speaker–utterance alignment, turn-taking logic, emotional and expressive alignment, and speaker-centric camera alignment.
In addition, MTAVG-Bench incorporates representative question formats, including multi-choice questions (single- and multiple-answer) for dimension-specific diagnostics and pairwise preference judgments for overall dialogue quality and prompt alignment, enabling fine-grained analysis beyond scalar scores.

In summary, our contributions are three-fold:
\begin{itemize}
\item We introduce MTAVG-Bench, a failure-driven diagnostic benchmark for multi-talker dialogue-centric audio-video generation, featuring high coverage and complexity through dialogue-centric prompts, multi-system T2AV synthesis, and fine-grained human annotations of multi-speaker failure cases.
\item MTAVG-Bench is organized into four progressive levels, namely signal fidelity, attribute consistency, social interaction, and cinematic expression, with fine-grained diagnostic dimensions and question-based protocols for both dimension-wise and overall assessment.
\item We provide new empirical insights into the limitations of current omni-modal evaluators and state-of-the-art T2AV systems in multi-speaker dialogue settings.
\end{itemize}

\begin{table*}[h]
\centering
\resizebox{\textwidth}{!}{
\begin{tabular}{lcccccccc}
\toprule
\textbf{Benchmarks} & 
\textbf{\#Video} & 
\textbf{\#QA} & 
\textbf{Dimen.} & 
\textbf{Failure-Mode} & 
\textbf{Modalities} &  
\textbf{Speaker-Centric}  &
\textbf{Multi-speaker}  &
\textbf{Dialogue}  \\
\midrule

Harmony-Bench~\cite{hu2025harmony} 
& 150 & -- & 3 & -- & T2AV & \xmark & \xmark & \xmark \\

JavisBench~\cite{Javisdit}   
& 10{,}140 & -- & 5 & -- & T2AV & \xmark & \xmark & \xmark \\

UniAVGen~\cite{Uniavgen}      
& 100 & -- & 3 & -- & T2AV & \xmark & \xmark & \xmark \\

VerseBench~\cite{wang2025universe}    
& 600 & -- & 4 & -- & T2AV & \xmark & \xmark & \xmark \\

VABench~\cite{hua2025vabench}       
& 1,300 & 14,300 & 15 & -- & I2AV/T2AV & \xmark & \xmark & \xmark \\

VideoHallu~\cite{li2025videohallu} 
& 120 & 3,233 & 4 & 13 & T2V & \xmark & \xmark & \xmark \\

Pistachio~\cite{li2025pistachio} 
& 4,962 & -- & 5 & 31 & T2V & \xmark & \xmark & \xmark \\

\midrule

\rowcolor{lightgray}
MTAVG-Bench (Ours) 
& 1,880 
& 2,410 
& 9 
& 37 
& T2AV 
& \cmark 
& \cmark 
& \cmark \\

\bottomrule
\end{tabular}

}
\vspace{-2ex}
\caption{Comparison of evaluation paradigms. Existing benchmarks mainly assess perceptual quality and alignment, while MTAVG-Bench additionally evaluates multi-speaker dialogue structure, tri-modal generation, and failure diagnosis.}
\label{tab:paradigm_compare}
\vspace{-1em}
\end{table*}

\section{Related Work}
\subsection{Speech-centric Audio-Visual Generation}
Traditional speech-centric audio-visual generation methods~\cite{zhou2021pose,prajwal2020lip,wang2023seeing,zhou2023learning} mainly focus on audio-driven talking-head synthesis and are typically limited to single-speaker and visually constrained settings. Recent works~\cite{wei2025mocha,multitalk,gan2025omniavatar,chen2025hunyuanvideo,ding2025kling} have extended this paradigm to multi-speaker scenarios, but still rely on given images and audio to synthesize multi-character videos.

With the emergence of commercial T2AV models~\cite{google2025veo3,openai2025sora2,alibaba2025wan25} trained on large-scale speech and sounding video data, movie-level multi-speaker dialogue generation from text prompts has become possible, where speech content, visual appearance, and multi-turn interactions are jointly synthesized. However, under this more structurally and interactionally complex setting, existing models~\cite{google2025veo3,openai2025sora2,alibaba2025wan25} still suffer from speaker identity inconsistency, incoherent turn-taking, and cross-modal semantic misalignment, which have not yet been systematically evaluated. Consequently, existing evaluation protocols fail to capture the structured understanding and interactional reasoning required for multi-speaker dialogue generation, motivating us to propose an understanding- and diagnosis-oriented benchmark for revealing hidden failure modes.

\subsection{Benchmarks for Audio-Visual Understanding and Generation}
Recent speech-related audio–visual understanding benchmarks such as AVUT~\cite{yang2025audio}, AV-SpeakerBench~\cite{nguyen2025see}, and AMUSE~\cite{chowdhury2025amuse} begin to incorporate speaker-centered and temporally grounded reasoning, but still fall short in modeling fine-grained speech semantics, multi-speaker interaction, and robust audio–visual grounding in complex dialogue scenes. On the other hand, MSU-Bench~\cite{wang2025msu} focus on speech-only multi-speaker understanding and do not capture visual–speaker alignment. Existing audio–visual generation benchmarks and evaluation frameworks~\cite{lan2025survey}, covering joint audio–video generation~\cite{Javisdit,wang2025universe}, comprehensive audio–visual generation~\cite{hua2025vabench}, primarily focus on the fidelity, synchronization, and semantic consistency of individual audio–video clips.
In contrast, we introduce MTAVG-Bench, the first benchmark for multi-talker audio–visual dialogue generation with fine-grained audio–visual error diagnosis and comprehensive evaluation in complex multi-speaker scenes.

\begin{table*}[t]
\centering
\small

% \vspace{-3ex}

\begin{tabular}{%
  p{0.18\textwidth}
  p{0.28\textwidth}
  p{0.48\textwidth}
}
\toprule
\textbf{Major Levels} & \textbf{Sub-dimensions} & \textbf{Evaluation Focus} \\
\midrule

\multirow{2}{=}{\textbf{Level 1: \\Signal Fidelity}}
& \textbf{Perceptual Video Quality (VQ)}
& Visual integrity of frames, including sharpness, temporal stability, and correct geometric rendering, \textit{without} flickering, blur, clipping, or missing body parts. \\
\cmidrule(lr){2-3}

& \textbf{Perceptual Speech Quality (SQ)}
& Acoustic integrity of speech, including continuity, cleanness, and naturalness, \textit{without} silence breaks, background noise, or artificial sound artifacts. \\
\midrule

\multirow{3}{=}{\textbf{Level 2: \\Attribute Consistency}}
& \textbf{Scene Consistency (SC)}
& Coherence of environment and setting across time, including location, time of day, and physical plausibility, \textit{without} unintended scene switches or violations of commonsense physics. \\
\cmidrule(lr){2-3}

& \textbf{Character Consistency (CC)}
& Stability of each speaker’s identity across time, including appearance, attributes, voice, and presence, \textit{without} visual distortion, voice drift, or identity mismatch. \\
\cmidrule(lr){2-3}

& \textbf{Audio-Visual Lip Synchronization (LS)}
& Temporal alignment between lip motion and speech audio, \textit{without} silent talking, speaking \textit{without} mouth movement, or mismatched lip–voice timing. \\
\midrule

\multirow{2}{=}{\textbf{Level 3: \\Social Interaction}}
& \textbf{Speaker-Utterance Alignment (SA)}
& Correct mapping between spoken content and speakers, including language, content, and attribution, \textit{without} narration shifts, wrong speaker assignment, or missing/extra utterances. \\
\cmidrule(lr){2-3}

& \textbf{Turn-Taking Logic (TT)}
& Temporal organization of dialogue turns, ensuring speakers \textit{do not} overlap, truncate, skip, or hallucinate turns, and that silence and transitions remain natural. \\
\midrule

\multirow{2}{=}{\textbf{Level 4: \\Cinematic Expression}}
& \textbf{Affective \& Expressive Alignment (EA)}
& Alignment between speech, emotion, and body behavior, ensuring natural gestures, prosody, and emotional reactions \textit{without} rigidity, flat tone, or mismatched actions. \\
\cmidrule(lr){2-3}

& \textbf{Speaker-Centric Camera Alignment (CA)}
& Camera framing and motion follow the active speaker and narrative intent, ensuring focus, tracking, and composition remain coherent with who is speaking. \\

\bottomrule
\end{tabular}

\vspace{-2ex}
\caption{Hierarchical evaluation taxonmomy of MTAVG-Bench. Four major levels encompassing nine sub-dimensions are systematically designed, inspired by common failure modes observed in popular T2AV models.}
\label{tab:evaluation_dimensions}
\end{table*}

\begin{figure}[t]
  \centering
  \includegraphics[width=\linewidth]{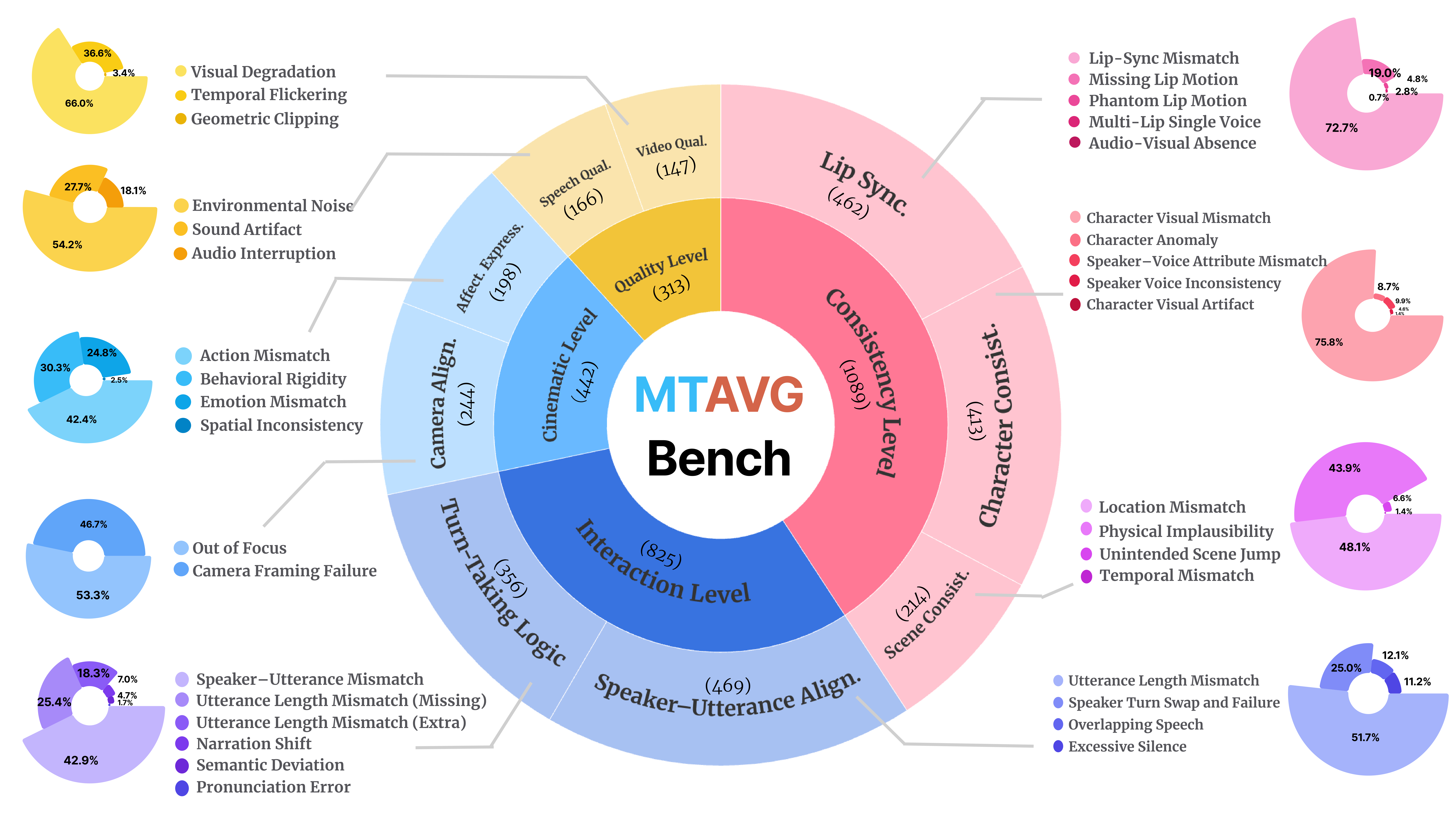}
  \vspace{-2ex}
  
  \caption{Data distribution of MTAVG-Bench.}
  \label{fig:distribution}
\end{figure}

\section{MTAVG-Bench}
As summarized in Table~\ref{tab:paradigm_compare}, MTAVG-Bench is a hierarchical diagnostic benchmark designed to evaluate multi-talker audio visual dialogue generation beyond surface realism. It comprises nine fine-grained metrics across four domains, Signal, Consistency, Interaction, and Cinematic Alignment, capturing identity persistence, temporal logic, and social dynamics to provide a holistic assessment of physical coherence and conversational fidelity in high-quality audio visual synthesis.

\begin{figure*}[t]
  \centering
  \includegraphics[width=\textwidth]{image/data.jpg}
  \vspace{-2ex}
  
  \caption{MTAVG-Bench construction and annotation pipeline. Multi-speaker dialogues are first rewritten by an LLM into structured prompts and used to generate multi-talker audio-visual clips with Veo 3.1, Wan 2.5, and Sora 2. The generated videos are analyzed to discover fine-grained failure cases, which are systematically mapped to a unified set of failure/evaluation dimensions. Based on this failure-dimension mapping, a failure-aware QA generator produces diverse evaluation questions that are further validated and refined by human experts.}
  \label{fig:data_construct}
\end{figure*}

\subsection{Evaluation Dimension Taxonomy}

To comprehensively assess the realism and coherence of multi-speaker AIGC-generated audio-visual content, we define four evaluation dimensions. As summarized in Table~\ref{tab:evaluation_dimensions}, these dimensions progress from low-level signal fidelity to high-level cinematic and expressive alignment, reflecting how human observers perceive realism in real-world conversational scenarios.

\subsubsection{Level 1: Signal Fidelity}

This level evaluates whether the generated audio and video streams are perceptually valid and free from low-level corruption. In multi-talker dialogue videos, signal artifacts such as visual flickering or audio glitches can disrupt all higher-level reasoning, making signal fidelity a fundamental prerequisite for meaningful evaluation. Signal Fidelity consists of two dimensions: \textit{Perceptual Video Quality}, which measures visual clarity, temporal stability, and geometric correctness of the frames, and \textit{Perceptual Speech Quality}, which assesses the acoustic continuity, cleanness, and naturalness of the generated speech. Failures at this level include blur, clipping, background noise, silence breaks, and artificial sound artifacts, all of which directly break the perceptual realism of the video.

\subsubsection{Level 2: Attribute Consistency}

This level examines whether the scene and the speakers remain stable and logically coherent across time. In multi-speaker settings, inconsistencies in environment, identity, or audio–visual correspondence often lead to confusion even when individual frames appear realistic.

Attribute Consistency includes three complementary dimensions. \textit{Scene Consistency} evaluates whether the location, lighting, time of day, and physical plausibility of the environment remain coherent without unintended switches or violations of commonsense physics. \textit{Speaker Consistency} measures the stability of each speaker’s visual appearance, attributes, voice, and presence across the video. \textit{Audio-Visual Lip Synchronization} assesses the temporal alignment between speech audio and lip motion, preventing silent talking, phantom speech, or lip–voice mismatch.

\subsubsection{Level 3: Social Interaction}

This level captures the core challenge of multi-talker dialogue generation: maintaining coherent conversational structure and correct speaker interactions over multiple turns. It consists of two dimensions. \textit{Speaker-Utterance Alignment} evaluates whether each spoken utterance is correctly attributed to the visible and active speaker, ensuring consistency between voice, character identity, and dialogue content. \textit{Turn-Taking Logic} measures whether speakers alternate naturally, without overlapping speech, abrupt truncation, skipped turns, or hallucinated participants. Errors at this level often produce videos that are perceptually realistic but socially incoherent.

\subsubsection{Level 4: Cinematic Expression}

This level evaluates whether the generated video achieves coherent cinematic and expressive presentation beyond correct dialogue. Even when speech and turn-taking are correct, failures in emotion, gesture, or camera control can severely degrade perceived realism. Cinematic Expression contains two dimensions. \textit{Affective and Expressive Alignment} assesses whether facial expressions, body movements, prosody, and emotional reactions are semantically aligned with the dialogue content. \textit{Speaker-Centric Camera Alignment} evaluates whether camera framing, focus, and motion follow the active speaker and narrative flow, ensuring that visual storytelling remains coherent with who is speaking.

\subsection{Benchmark Construction}

\subsubsection{Data Pipeline}

\textbf{Audio-Video Generation.} As illustrated in Figure~\ref{fig:data_construct}, the dataset is built through a structured annotation pipeline. We first generate a large set of multi-turn dialogue prompts and feed them into a text-to-audio-video synthesis system to produce multi-speaker dialogue videos. An agent-based filtering mechanism is then applied to automatically discard videos without apparent errors, ensuring that the dataset focuses on samples containing at least one observable failure.

\noindent\textbf{Failure Case Annotation.} In the failure discovery stage, each generated video is first processed by automated agents to identify potential abnormal or unnatural behaviors, after which human annotators carefully review the candidates to confirm true failures. Each confirmed failure is then manually annotated and mapped to one or more dimensions of the evaluation framework through a failure-to-dimension alignment process. For example, a single video may be labeled as exhibiting both a lip synchronization error under the Signal Fidelity dimension and a turn-taking logic failure under the Social Interaction dimension. Human verification and refinement are applied throughout this stage to ensure that every annotated failure is both perceptually grounded and semantically consistent with the underlying video content.

\noindent\textbf{Question-Answer Pairs Generation.} After failures are identified and categorized, we employ an LLM-assisted procedure to generate diagnostic question-answer (QA) pairs. For each failure instance, a corresponding QA item is created, consisting of a question explicitly targeting the failure (e.g., whether turn-taking is correct) and a set of candidate answers. Depending on the failure type, questions are formulated as single-choice, multiple-choice, or pairwise-comparison items. Importantly, all QA items are grounded in real failures identified by human annotators in multi-speaker audio-video dialogues, rather than hypothetical cases, allowing the benchmark to faithfully reflect both generator performance in multi-speaker, multi-turn settings and the ability of multimodal models to recognize and diagnose such failures. Each failure-conditioned video and its corresponding QA pairs are independently annotated by three annotators randomly sampled from a 21-person pool over a verification process lasting more than one month. Cases with complete disagreement are adjudicated by an expert panel to establish gold labels, and each failure-to-dimension mapping and QA pair undergoes an additional round of double review to ensure tight alignment with the video content. Through this semi-automated process, every video containing a failure yields at least one high-quality diagnostic QA pair for model evaluation.

\subsubsection{Data Distribution and Coverage}
We organize MTAVG-Bench around 37 fine-grained failure modes spanning 9 evaluation dimensions, yielding thousands of human-annotated failure cases, each paired with a diagnostic question. Figure~\ref{fig:distribution} shows their distribution across the nine dimensions. Lip synchronization errors are the most frequent, highlighting that in multi-speaker dialogue generation, lip-sync consistency is a structural prerequisite for stable speaker attribution and dialogue grounding rather than merely low-level audio-video alignment. Turn-taking failures follow, reflecting challenges in maintaining coherent multi-turn dialogue, while speaker-utterance mismatches rank third, revealing persistent difficulty in preserving speaker identity and voice consistency. Notably, some cases exhibit plausible lip motion while still violating higher-level constraints, such as when spoken content deviates from the text prompt despite temporal alignment. This suggests that audio-video synchronization alone is insufficient to ensure dialogue-level consistency in complex multi-speaker settings.

\begin{table*}[t]
\centering
\footnotesize
\setlength{\tabcolsep}{0pt}
\begin{tabular*}{\textwidth}{@{\extracolsep{\fill}}l c cc ccc cc cc c@{}}
\toprule
\multirow{2}{*}{\textbf{Model}} & \multirow{2}{*}{\textbf{Size}} 
& \multicolumn{2}{c}{\textbf{Signal Level}} 
& \multicolumn{3}{c}{\textbf{Consistency Level}} 
& \multicolumn{2}{c}{\textbf{Interaction Level}} 
& \multicolumn{2}{c}{\textbf{Cinematic Level}} 
&\multirow{2}{*}{\textbf{Avg.}} \\
\cmidrule(lr){3-4} \cmidrule(lr){5-7} \cmidrule(lr){8-9} \cmidrule(lr){10-11}

 & 
   & \textbf{VQ} & \textbf{SQ}
   & \textbf{SC} & \textbf{CC} & \textbf{LS}
   & \textbf{SA} & \textbf{TT}
   & \textbf{EA} & \textbf{CA}
   &  \\
\midrule

\multicolumn{12}{l}{\textbf{Proprietary Omni Models}} \\
\texttt{Gemini 2.5 Flash} & -- & 48.40 & 51.66 & 47.06 & 30.53 & 61.21 & 56.87 & 53.83 & 45.34 & \underline{52.87} & 49.75 \\
\texttt{Gemini 2.5 Flash Thinking} & -- & 57.60 & 51.99 & 44.61 & 29.02 & \underline{64.36} & 56.38 & 55.74 & 45.34 & 48.36 & 50.38 \\
\texttt{Gemini 2.5 Pro Thinking} & -- & \underline{58.40} & 49.67 & \underline{50.74} & 38.69 & \textbf{65.62} & \underline{65.61} & 55.99 & \underline{49.74} & \textbf{53.28} & \underline{54.19} \\
\texttt{Gemini 3 Flash}  & -- & 51.20 & 52.32 & 46.08 & 40.76 & 63.22 & \textbf{68.78} & \underline{56.39} & 49.22 & 49.18 & 53.02 \\
\texttt{Gemini 3 Pro} & -- & \textbf{70.40} & \underline{55.30} & \textbf{53.43} & 41.19 & 52.90 & 68.63 & \textbf{60.83} & \textbf{58.03} & 50.82 & \textbf{56.84} \\

\midrule
\multicolumn{12}{l}{\textbf{Open-sourced Omni Models}} \\
\texttt{Ming-Omni} & 30B & 40.80 & 39.40 & 33.33 & 29.37 & 51.89 & 41.09 & 37.51 & 36.27 & 45.08 & 39.42 \\
\texttt{Video-Salmonn2-Plus} & 7B & 43.21 & 42.57 & 41.46 & 33.64 & 54.48 & 37.16 & 37.51 & 36.02 & 44.26 & 41.15 \\
\texttt{MiniCPM-o 2.6} & 7B & 41.20 & 47.35 & 28.68 & 34.76 & 52.90 & 45.18 & 37.81 & 35.49 & 52.05 & 41.71 \\
\texttt{Qwen2.5-Omni} & 7B & 44.00 & 42.38 & 36.76 & \underline{44.65} & 60.45 & 36.32 & 40.89 & 38.86 & 40.98 & 42.81 \\
\texttt{Qwen3-Omni} & 30B & 52.00 & 47.35 & 39.95 & 37.69 & 38.79 & 51.01 & 46.12 & 46.89 & 51.64 & 45.72 \\
\texttt{Video-LLaMA2} & 7B & 48.80 & 50.00 & 48.04 & \textbf{45.12} & 47.48 & 50.88 & 39.78 & 45.85 & 51.23 & 47.46 \\
\texttt{Ola-Omni} & 7B & 46.40 & \textbf{55.96} & 37.50 & 36.79 & 61.96 & 52.24 & 43.25 & 46.11 & 50.00 & 47.80 \\

\bottomrule
\end{tabular*}
\vspace{-2ex}
\caption{Evaluation results on MTAVG-Bench. We report performance across four hierarchical levels, encompassing nice distinct dimensions. \textbf{Avg.} represents the cumulative mean across all metrics. The best and second-best results are highlighted in \textbf{bold} and \underline{underline}.}

\label{tab:main}
\end{table*}

\begin{table*}[t]
\centering
\small
\setlength{\tabcolsep}{2.5pt}
\begin{tabular*}{\textwidth}{@{\extracolsep{\fill}}l cc ccc cc cc cc@{}}
\toprule

\multirow{2}{*}{\textbf{Eval. Setting}} 
& \multicolumn{2}{c}{\textbf{Signal Level}} 
& \multicolumn{3}{c}{\textbf{Consistency Level}} 
& \multicolumn{2}{c}{\textbf{Interaction Level}} 
& \multicolumn{2}{c}{\textbf{Cinematic Level}} 
& \multirow{2}{*}{\textbf{Avg.}} \\

\cmidrule(lr){2-3} 
\cmidrule(lr){4-6} 
\cmidrule(lr){7-8} 
\cmidrule(lr){9-10}

& \textbf{VQ}
& \textbf{SQ}
& \textbf{SC}
& \textbf{CC}
& \textbf{LS}
& \textbf{SA}
& \textbf{TT}
& \textbf{EA}
& \textbf{CA}
\\
\midrule

\texttt{Gemini 3 Pro (Full)} 
& 70.40 
& 55.30 
& 53.43 
& 41.19 
& 52.90 
& 68.63 
& 60.83 
& 58.03 
& 50.82 
& 56.84 \\

\texttt{- Without Audio input}
& 63.20 
& 50.33 
& 51.47 
& 33.85 
& 37.78 
& 48.55 
& 44.61 
& 49.74 
& 51.64 
& 47.46 \\

\texttt{- Without Gen. Prompt Align.} 
& 60.80 
& 52.98 
& 45.59 
& 37.16 
& 56.68 
& 56.71 
& 55.69 
& 52.85 
& 48.77 
& 51.36 \\

\bottomrule
\end{tabular*}

\vspace{-2ex}
\caption{Ablation study on input conditions used for assessing multi-talker video generation.}
\vspace{-2ex}
\label{tab:ablation}

\end{table*}

\section{Experiments}
\subsection{Experiment Setup}
\textbf{Models.} We evaluate a broad set of state-of-the-art omni-modal models with native audio--video understanding for multi-talker dialogue. Our benchmark covers both proprietary and open-source systems across diverse architectures and training paradigms. For proprietary models, we evaluate the Gemini family~\cite{team2023gemini}, including Gemini 3 Pro, Gemini 3 Flash, Gemini 2.5 Pro (Thinking), Gemini 2.5 Flash (Thinking), and Gemini 2.5 Flash, all of which support end-to-end audio-video perception and multimodal reasoning. For open-source models, we include representative publicly available omni with native audio--video input support: Video-LLaMA2~\cite{cheng2024videollama}, MiniCPM-o 2.6~\cite{yu2025minicpm}, Ola~\cite{liu2025ola}, Qwen2.5-Omni~\cite{xu2025qwen2}, Video-Salmonn2-Plus~\cite{tang2025video}, Qwen3-Omni~\cite{xu2025qwen3omnitechnicalreport}, and Ming-Omni~\cite{ai2025ming}. These models range from lightweight 7B-scale systems to larger 30B-scale omni architectures, enabling systematic comparison across model sizes and design choices.

\subsection{Evaluation Protocol and Metrics}
We evaluate each model using a hierarchical, failure-driven protocol for multitalker audio--visual dialogue understanding. Each generated clip is assessed across nine fine-grained failure dimensions organized into four levels: signal fidelity (VQ, SQ), attribute consistency (SC, CC, LS), social interaction (SA, TT), and cinematic alignment (EA, CA). For each dimension, failure-aware questions are constructed in three formats---single-answer multiple-choice question (MCQ), multiple-answer MCQ, and pairwise comparison---to probe the model's ability to detect specific generation errors.

\paragraph{Per-question scoring.}
Let question $i$ have ground-truth answer(s) $G_i$ and model prediction $P_i$. Each question receives a normalized score $s_i \in [0,1]$ defined as:
\begin{equation}
s_i =
\begin{cases}
\mathbb{I}[P_i = G_i], & \text{(single-choice MCQ)} \\[4pt]
\dfrac{|P_i \cap G_i|}{|G_i|}, & \text{(multiple-choice MCQ)} \\[8pt]
\mathbb{I}[P_i = G_i], & \text{(pairwise comparison)},
\end{cases}
\end{equation}
where $\mathbb{I}[\cdot]$ denotes the indicator function.

\paragraph{Dimension-wise and overall scores.}
For each failure dimension $d \in \mathcal{D}$, let $\mathcal{Q}_d$ be the set of associated questions. The dimension-level score is computed by:
\begin{equation}
\text{Score}_d = \frac{1}{|\mathcal{Q}_d|} \sum_{i \in \mathcal{Q}_d} s_i .
\end{equation}
The overall performance is reported as the unweighted mean across all dimensions:
\begin{equation}
\text{Avg.} = \frac{1}{|\mathcal{D}|} \sum_{d \in \mathcal{D}} \text{Score}_d .
\end{equation}
The values in Table~\ref{tab:main} correspond to $\text{Score}_d$, and \textbf{Avg.} is computed accordingly.

\begin{table*}[t]
\centering
\small
\setlength{\tabcolsep}{2.5pt}
\begin{tabular*}{\textwidth}{@{\extracolsep{\fill}}l cc ccc cc cc cc@{}}
\toprule
\multirow{2}{*}{\textbf{Gen. Models}} 
& \multicolumn{2}{c}{\textbf{Signal Level}} 
& \multicolumn{3}{c}{\textbf{Consistency Level}} 
& \multicolumn{2}{c}{\textbf{Interaction Level}} 
& \multicolumn{2}{c}{\textbf{Cinematic Level}} 
& \multirow{2}{*}{\textbf{Avg.}} \\
\cmidrule(lr){2-3} \cmidrule(lr){4-6} \cmidrule(lr){7-8} \cmidrule(lr){9-10}
& \textbf{VQ}
& \textbf{SQ}
& \textbf{SC}
& \textbf{CC}
& \textbf{LS}
& \textbf{SA}
& \textbf{TT}
& \textbf{EA}
& \textbf{CA}
\\
\midrule
\multicolumn{11}{l}{\textit{Commercial models}} \\
\midrule

\texttt{Sora2}
& 85\% 
& \textbf{91\%} 
& \textbf{82\%} 
& \textbf{67\%} 
& \textbf{79\%} 
& \textbf{75\%} 
& 64\% 
& \textbf{76\%} 
& \textbf{41\%} 
& \textbf{73\%} \\

\texttt{VEO 3.1} 
& 92\% 
& 90\% 
& 48\% 
& 59\% 
& 73\% 
& 61\% 
& \textbf{72\%} 
& 70\% 
& 40\% 
& 67\% \\

\texttt{WAN 2.5}
& \textbf{95\%} 
& 83\% 
& 54\% 
& 51\% 
& 46\% 
& 41\% 
& 60\% 
& 64\% 
& 30\% 
& 58\% \\

\midrule
\multicolumn{11}{l}{\textit{Open-source models}} \\
\midrule

\texttt{LTX 2.3} & 74\% & 78\% & 38\% & 43\% & 58\% & 24\% & 19\% & 33\% & 22\% & 43\% \\
\texttt{Ovi}     & 68\% & 66\% & 29\% & 34\% & 27\% & 18\% & 12\% & 20\% & 21\% & 33\% \\
\bottomrule
\end{tabular*}
\vspace{-2ex}
\caption{Human evaluation of multitalker T2AV models, revealing a large gap between perceptual quality and multi-speaker consistency, interaction, and camera alignment.}
\label{tab:human_eval}
\vspace{-1ex}
\end{table*}

\subsection{Benchmark Results}
As shown in Table~\ref{tab:main}, Gemini 3 Pro achieves the strongest overall performance on MTAVG-Bench and shows clear advantages on higher-level interaction and cinematic dimensions, including speaker alignment (SA), turn-taking (TT), and expression alignment (EA). This suggests that it not only perceives audio--visual signals accurately, but also demonstrates stronger capability in evaluating complex cross-speaker and cross-modal behaviors. Notably, performance differences are driven primarily by interaction modeling rather than signal-level fidelity. Although many models perform similarly on signal-level metrics, much larger gaps emerge on interaction dimensions, where Gemini 3 Pro and Gemini 3 Flash outperform most open-source models by substantial margins on speaker alignment and turn-taking. This indicates that many models still struggle to determine who is speaking and when, leading them to judge conversationally incorrect videos too favorably. Although Qwen3-Omni and Ming-Omni both have 30B parameters, they are still outperformed by the 7B Ola model on several speech and interaction metrics. Overall, these results suggest that reliable diagnosis of video failure modes in multi-speaker T2AV settings depends not only on audio--visual perception, but also on multimodal models having a stronger ability to understand and model high-level interaction structure, speaker attribution, and turn-taking logic.

\begin{table}[t]
\centering
\small
\setlength{\tabcolsep}{6pt}
\begin{tabular}{lccc}
\toprule
\textbf{Model} & \textbf{Sora} & \textbf{Veo} & \textbf{Wan} \\
\midrule
\texttt{Gemini 3 Pro}                & 55.3 & 58.9 & 55.5 \\
\texttt{Gemini 2.5 Pro Thinking}     & 53.4 & 54.5 & 51.7 \\
\texttt{Qwen3-Omni}                  & 44.2 & 49.3 & 44.5 \\
\texttt{Ola-Omni}                    & 46.1 & 49.4 & 47.3 \\
\bottomrule
\end{tabular}
\vspace{-2ex}
\caption{Evaluator performance (\%) on single-generator subsets of MTAVG-Bench. Although absolute scores vary across Sora, Veo, and Wan subsets, model rankings remain largely stable, indicating limited dependence on generator-specific artifact distributions.}
\label{tab:generator_ablation}

\end{table}

\section{Further Analysis}
\subsection{Ablation Study on Input Conditions}
Table~\ref{tab:ablation} reports an ablation study on Gemini 3 Pro, highlighting the importance of audio input and generative prompt alignment for diagnosing failures in multitalker audio–visual dialogue generation. Removing audio causes the largest performance drop, especially on interaction metrics, where speaker attribution and turn-taking fall from 68.63 to 48.55 and from 60.83 to 44.61, respectively, demonstrating the necessity of speech for tracking speakers and dialogue structure. Disabling prompt alignment also degrades performance, mainly on consistency and cinematic metrics such as character identity and camera alignment, due to weakened semantic grounding between intended dialogue and generated video. Overall, reliable failure diagnosis requires both audio-based interaction modeling and prompt-aware semantic grounding for multimodal and narrative coherence. %narrative and cinematic coherence

\subsection{Overall Quality Estimation}

MTAVG-Bench is designed as a diagnostic benchmark for structural failure reasoning in multi-talker T2AV generation. To assess whether the failure patterns highlighted by the benchmark also arise under the natural output distribution of current generators, we additionally estimate overall quality on the full set of unfiltered samples from Sora, Veo, and Wan, rather than restricting evaluation to curated failure cases. As shown in Table~\ref{tab:natural_dist_eval}, modern generators maintain relatively strong perceptual or signal-level quality, but exhibit substantially weaker performance on higher-level dialogue-centric dimensions such as speaker consistency, social interaction, turn-taking, emotional alignment, and camera alignment. This trend is consistent across generators. These results suggest that the failure patterns emphasized in MTAVG-Bench are not artifacts of failure mining, but recurrent weaknesses that also emerge under the natural generation distribution. Therefore, the benchmark captures systematic structural deficiencies of current multi-talker T2AV systems rather than isolated corner cases.

\begin{table*}[t]
\centering
\small
\setlength{\tabcolsep}{5pt}
\begin{tabular*}{\textwidth}{@{\extracolsep{\fill}}l c ccc cccccc@{}}
\toprule
\textbf{Generator}
& \textbf{Speech Q.}
& \textbf{T-V}
& \textbf{T-A}
& \textbf{A-V}
& \textbf{SC}
& \textbf{CC}
& \textbf{SA}
& \textbf{TT}
& \textbf{EA}
& \textbf{CA} \\
\midrule

\multicolumn{11}{l}{\textit{Commercial models}} \\
\cmidrule(lr){1-11}
\texttt{Sora2}   & 3.14 & 0.33 & 0.22 & 0.29 & 4.87 & 4.76 & 4.91 & 4.85 & 4.66 & 4.94 \\
\texttt{VEO 3.1} & 3.36 & 0.36 & 0.19 & 0.35 & 4.10 & 4.62 & 4.81 & 4.51 & 4.30 & 4.83 \\
\texttt{WAN 2.5} & 2.68 & 0.35 & 0.16 & 0.20 & 4.13 & 4.54 & 4.66 & 4.25 & 3.91 & 4.72 \\

\midrule
\multicolumn{11}{l}{\textit{Open-source models}} \\
\cmidrule(lr){1-11}
\texttt{LTX 2.3} & 3.18 & 0.26 & 0.17 & 0.18 & 3.92 & 4.27 & 4.48 & 3.98 & 3.83 & 4.68 \\
\texttt{Ovi}     & 2.77 & 0.25 & 0.19 & 0.15 & 3.76 & 3.94 & 3.63 & 3.44 & 3.52 & 4.53 \\

\bottomrule
\end{tabular*}
\vspace{-2ex}
\caption{Overall quality estimation under the natural output distribution of three T2AV generators on unfiltered multi-talker dialogue videos. While perceptual quality and basic cross-modal correspondence remain relatively strong, higher-level structural dimensions such as consistency, interaction, and cinematic alignment remain substantially more challenging.}
\label{tab:natural_dist_eval}
\end{table*}

\subsection{Ablation on Generator-Specific Bias}

To evaluate whether evaluator rankings depend heavily on the specific T2AV generators used to construct the benchmark, we partition the evaluation set by generator and measure evaluator performance separately on the Sora-only, Veo-only, and Wan-only subsets. Table~\ref{tab:generator_ablation} shows that, although the absolute scores vary slightly across generators, the relative ranking of evaluators remains largely stable. In particular, the strongest models consistently outperform weaker ones across all three generator-specific subsets. This indicates that evaluator performance is driven more by high-level semantic reasoning and structural failure identification than by overfitting to generator-specific artifact distributions. Thus, the ranking conclusions reported in the main paper are robust to the choice of underlying T2AV generator.

\begin{figure}[t]
  \centering
  \includegraphics[width=\linewidth]{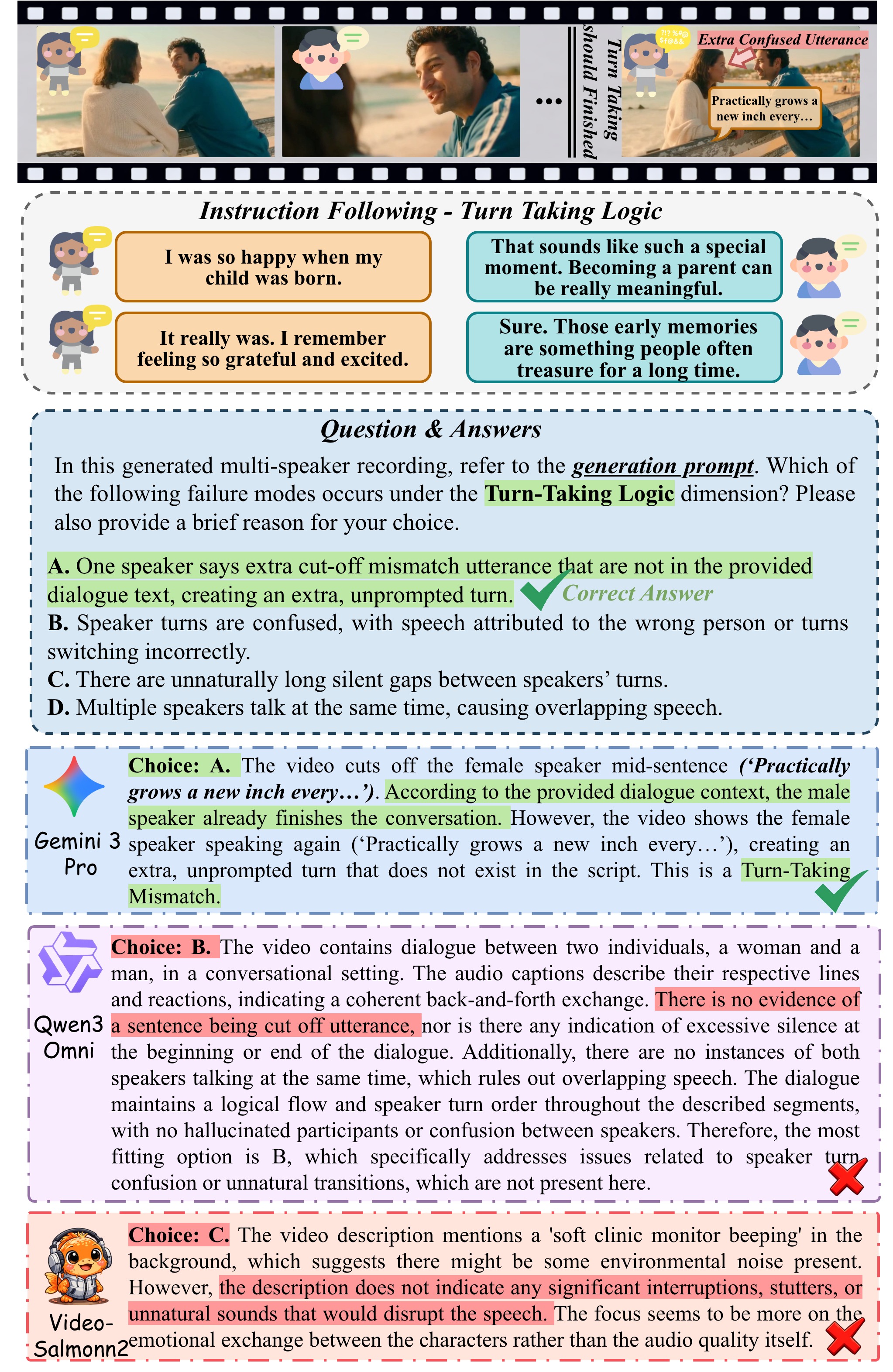}
  \vspace{-3ex}
  \caption{Qualitative Result for Instruction-Following under the Turn-Taking Logic Dimension.}
  \label{fig:case_study}
  \vspace{-2ex}
\end{figure}

\subsection{Human Success Rates on T2AV Models}
Table~\ref{tab:human_eval} presents human-judged success rates on a randomly sampled set of 3,000 multitalker T2AV videos prior to failure mining, with each generation model generating 1,000 videos from the same prompt set. Twenty human annotators assess whether each evaluation dimension is satisfied, reflecting perceptual success from a human perspective. Although these strong generation models achieve high performance on signal-level quality (VQ and SQ), their accuracy degrades markedly on multi-speaker consistency, interaction, and cinematic alignment, with frequent errors in character consistency, turn-taking, speaker alignment, and camera focus. VEO 3.1 achieves the best turn-taking performance, while Wan 2.5 leads in visual quality. 
Sora2 is the strongest overall model, yet it attains less than 42\% success on camera alignment, highlighting a substantial gap between perceptual realism and coherent multi-speaker storytelling. 
These findings indicate that current T2AV models prioritize visual and acoustic fidelity at the expense of social structure, speaker identity, and narrative focus, which remain the principal bottlenecks in multitalker audio-visual generation.

\subsection{Case Study}
To illustrate the diagnostic power of MTAVG-Bench, Figure~\ref{fig:case_study} presents a representative failure case from the Turn-Taking Logic (TT) dimension. Although the T2AV model generates a visually natural, high-fidelity video, it exhibits a critical structural error: the female speaker continues speaking after the scripted dialogue ends, producing an extra unprompted utterance. Video-Salmonn2~\cite{tang2025video} fails to reason about the dialogue structure and instead focuses on low-level acoustic cues such as background noise and beeps, reflecting a bias toward generic audio-visual events rather than communicative logic. Qwen3-Omni~\cite{xu2025qwen3omnitechnicalreport} attempts to analyze the interaction flow but exhibits cognitive hallucination, incorrectly judging the dialogue as logically consistent and overlooking the extra turn. In contrast, Gemini 3 Pro~\cite{google2025gemini3} correctly identifies the discrepancy between the script and the generated video, explicitly detecting the unprompted continuation. These results show that MTAVG-Bench goes beyond coarse quality assessment by exposing fine-grained failures in social interaction and dialogue structure, providing a rigorous framework for diagnosing high-level coherence in T2AV systems.

\section{Conclusion}

In this paper, we introduced MTAVG-Bench, a failure-driven diagnostic benchmark for multi-talker text-to-audio-video dialogue generation. Through a four-level evaluation framework covering signal fidelity, consistency, interaction, and cinematic expression, MTAVG-Bench enables fine-grained diagnosis of structural failure modes beyond existing benchmarks. Experiments on proprietary and open-source omni-models demonstrate the difficulty of reliable failure identification in multi-speaker T2AV outputs, while analyses of state-of-the-art generators reveal persistent weaknesses in speaker identity, turn-taking, and audio-visual grounding. We hope MTAVG-Bench will facilitate more rigorous evaluation and support the development of more reliable and controllable multi-talker audio-visual generation systems.

\section*{Limitation}

The proposed MTAVG-Bench is a valuable resource for evaluating synthesized audio-visual multi-talker videos and has the potential to support the future development of multimodal large language models and video generation systems.
Below, we provide further discussion on the limitations of this work.
In our setting, the performance of the proposed benchmark and evaluation framework may be influenced by two factors. \textit{First}, the multimodal representation capability of large multimodal models varies substantially across audio and visual modalities, and their input processing strategies are not uniform. This is particularly critical in talking-face and multitalker dialogue scenarios, where accurate failure diagnosis requires high-frequency alignment between speech and facial motion. However, current multimodal models have rarely been exposed to such fine-grained audio-visual synchronization errors during training, making this a largely out-of-distribution (OOD) problem. While supervised fine-tuning and multimodal alignment strategies provide promising directions, reliable detection of these failure modes remains challenging. \textit{Second}, the stochastic “sampling” nature of generative video models and their differing inductive biases lead to highly imbalanced distributions of failure modes across models. Different generators tend to exhibit distinct strengths and weaknesses, resulting in uneven coverage of error types in the generated data. This imbalance complicates downstream post-training of evaluators, such as SFT or RL, which require more uniformly distributed supervision across failure categories.
Together, these two factors highlight challenges in both diagnosing and learning from failure modes in multitalker audio-visual generation.

\section*{Ethics Statement}

This study follows the basic principles of responsible academic research. All data used in this work are obtained from publicly available sources and do not contain personal or sensitive information. The textual inputs are primarily drawn from open-source datasets and other publicly available text resources, and are used solely for academic research purposes. The audio-video data generated from these texts are synthetic outputs produced by models; they do not represent the authors' personal views and are not intended to influence, promote, or endorse any particular opinion. The generated audio-video content is used only for academic research and method validation, and not for impersonation, deceptive communication, misinformation, or other misleading applications. Since the source texts are publicly available, the outputs of our text-to-audio-video generation models should not be interpreted as expressing the views of any particular group, culture, or social standpoint. We emphasize that the relevant generative technologies are developed and used within lawful, compliant, and responsible frameworks. In addition, all datasets and models used in this study are permitted for academic research and comply with their respective licensing requirements.

\section*{Acknowledgments}
This work has been supported by Natural Science Foundation of Fujian Province (2025J01297). It was also supported by Inkeverse Group Limited. The authors would like to thank Jiahao Pan, Yukun Su, Xinhui Wang, Jiahui Wu, Duo Huang, Qiongxuan Wu, Mingjie Tan, Yanglin Zhang, Guoyuan Liu, Manning Luo, Jiarui Zhan, Qiqi Zhao, and other friends for their valuable discussions and contributions to the annotation work supporting this research.

\newpage

\bibliography{reference}

\newpage
\appendix

%\section{Supplementary}
%\label{sec:appendix}

\clearpage
\section{Prompt Design for Benchmark Construction and Evaluation}

We designed a series of structured system prompts to ensure high-fidelity video generation and rigorous model evaluation. These prompts explicitly specify the roles, constraints, and output formats for the different models involved in the pipeline. To construct diverse multi-speaker dialogue scenarios, we drew on the EmpatheticDialogues~\cite{rashkin2019towards} dataset and adapted its dialogue materials into structured JSON prompts for text-to-audio-video generation. Our data construction spans 32 emotion-driven dialogue themes and hundreds of randomized character–scene combinations, resulting in over 3,000 generated videos for large-scale failure mining. Human review was then conducted to identify representative failure cases. Importantly, these failures arise naturally from diverse script-to-T2AV generation with advanced models, rather than from manually injected perturbations.

\subsection{Video Generation Prompt}
To synthesize movie-level multi-talker videos from textual descriptions, we employ a ``Audio-Video Generation Prompt'' system instruction to generate better generation prompt for multi-talker dialogue-centric audio-video generation (as shown in Figure~\ref{fig:prompt_design_generating}).

\subsection{Evaluation Prompts}
For the evaluation phase, we position the VLM as a ``Senior Diagnostic Auditor.'' We designed two distinct prompt templates:
\begin{enumerate}
    \item \textbf{Single Choice Inference (Figure~\ref{fig:prompt_design_inference}):} Used for dimension-specific diagnostics. It requires the model to output a strictly formatted JSON object containing step-by-step forensic reasoning and the final choice.
    \item \textbf{Diagnostic Specialist (Figure~\ref{fig:prompt_design_specialist}):} A general template for identifying specific failures. It forces the model to cite specific visual or auditory evidence (e.g., ``Speaker A's mouth is closed while voice is heard'') before drawing conclusions.
\end{enumerate}

\section{Additional Qualitative Case Studies}
\label{subsec:case_studies}

We present more qualitative examples to demonstrate how MTAVG-Bench distinguishes model capabilities across different granularity levels.

\paragraph{Interaction Level: Speaker--Utterance Alignment.}
Figure~\ref{fig:case-3} presents a failure case in which the generated video contains an unprompted extra utterance (``\textit{Doesn't make it any less humiliating}'') spoken by the male character, despite the script explicitly requiring silence at that moment. Gemini~3~Pro correctly classifies this error as an \emph{Utterance Length Mismatch} (Choice~A), demonstrating strong instruction-following and fine-grained alignment capabilities. In contrast, Qwen3~Omni and Video-Salmonn2 misidentify the issue as either a speaker attribution error or a general inconsistency, failing to capture the true nature of the violation. This illustrates the difficulty of diagnosing subtle interaction-level errors in multimodal generation.

\paragraph{Cinematic Level: Camera Alignment.}
Figure~\ref{fig:case-4} compares two videos generated from the same prompt. Video~A exhibits professional cinematic conventions, such as over-the-shoulder framing and consistent spatial blocking, while Video~B shows unnatural ``breaking the fourth wall'' behavior, with characters staring directly into the camera. Gemini~3~Pro correctly prefers Video~A and supports its choice with concrete cinematic cues (e.g., ``teal and orange contrast'' and ``depth of field''), indicating that the benchmark effectively evaluates high-level aesthetic quality and narrative coherence in generated videos.

\section{Data Distribution Analysis}
\label{subsec:data_stats}

\begin{figure*}[t]
  \centering
  \includegraphics[width=\textwidth]{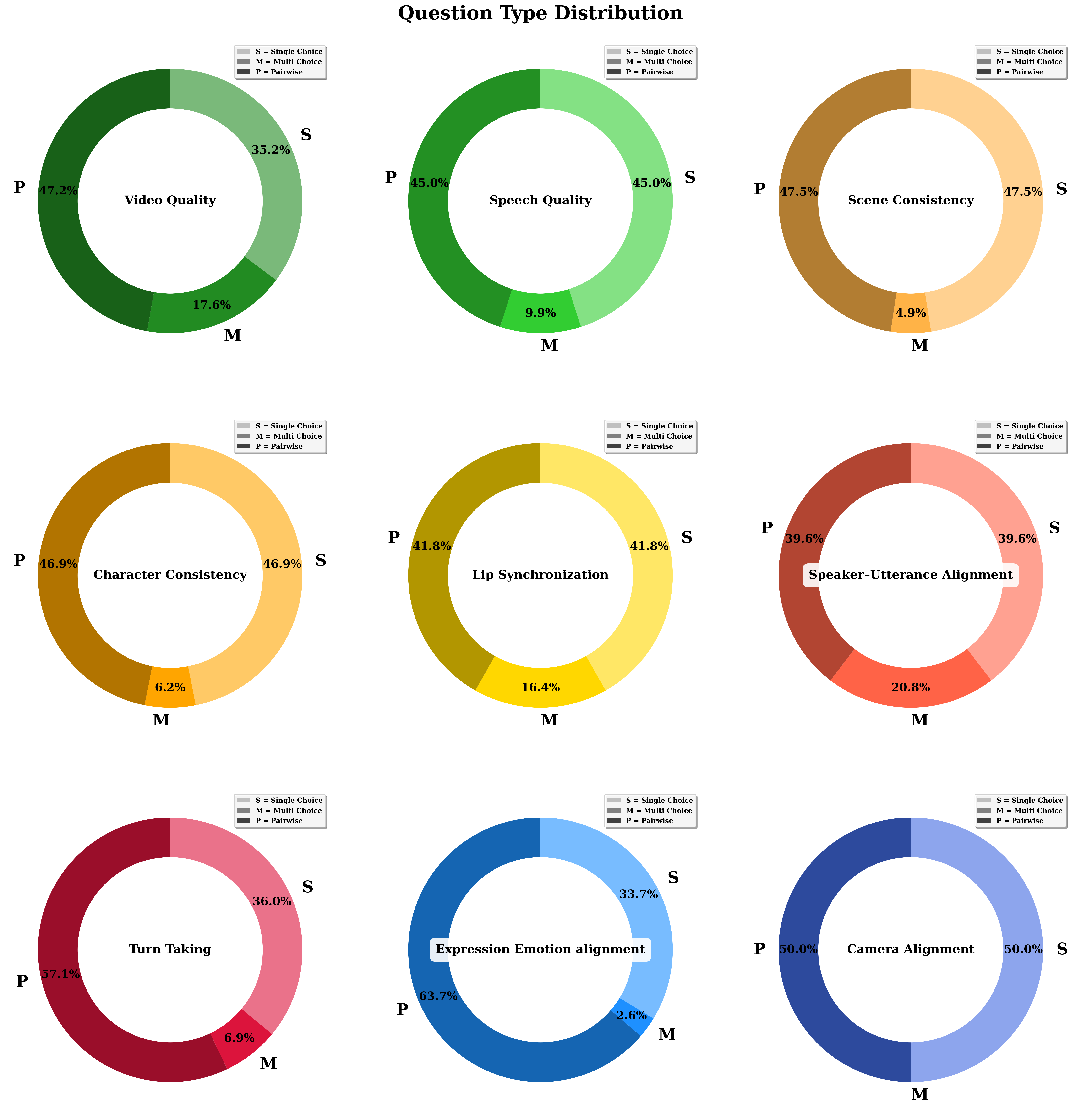}
  \caption{Breakdown of question types for each evaluation sub-dimension in MTAVG-Bench.}
  \label{fig:question_dist}
\end{figure*}

Figure~\ref{fig:question_dist} provides a breakdown of question types across the high-level evaluation dimensions. Within the \textbf{Consistency Level}, dimensions such as Scene Consistency and Character Consistency exhibit a highly symmetrical distribution between Single-Choice and Pairwise formats (approximately 47\% each), reflecting the dual objective of identifying absolute stability errors while also enabling relative model comparisons. The \textbf{Interaction Level} adopts a more task-dependent strategy: Turn Taking relies heavily on Pairwise comparisons (57.1\%) to evaluate conversational flow, whereas Speaker--Utterance Alignment incorporates a notable proportion of Multi-Choice questions (20.8\%) to diagnose more complex synchronization failures. Finally, the \textbf{Cinematic Level}, represented by Camera Alignment, focuses on evaluating whether a designated camera angle is audiovisually consistent with the scene.

\section{Annotation Quality and Reliability}
\label{subsec:annotation_quality}

To ensure annotation quality, we adopt a multi-stage protocol with redundancy and expert adjudication. Our annotation team consists of 21 annotators with master's- or doctoral-level training, who were organized into rotating groups for independent labeling and cross-verification over a period of more than one month. As illustrated in Figure~\ref{fig:data_construct}, the overall workflow combines rule-based annotation, multi-annotator review, and expert refinement to ensure both consistency and reliability. We use Label Studio as the annotation platform throughout the entire process. In the first stage, each generated video is manually diagnosed at the sample level with respect to nine evaluation dimensions. For each video, annotators identify failure modes under the corresponding dimensions and also assess the success states of the same nine dimensions, allowing us to capture both error patterns and successful generation signals in a unified framework. Each failure-conditioned sample and its corresponding labels are independently annotated by three annotators randomly selected from the pool of 21.

Quality control is enforced at both the failure labeling and QA construction stages. For failure annotation, cases with complete disagreement are resolved through expert adjudication to establish gold-standard labels. For QA construction, we again use Label Studio to refine the generated QA pairs based on the reference video, the original prompt, and the candidate question-answer pairs. Following the workflow in Figure~\ref{fig:data_construct}, annotators first review answer candidates independently under shared annotation rules; if the candidate answers are consistent, the QA pair is accepted directly, while inconsistent cases are further discussed and refined by an expert annotator to produce the final validated question-answer pair. Failure-to-dimension mappings and QA pairs therefore undergo structured multi-stage verification before inclusion in the benchmark.

We measure inter-annotator agreement (IAA) using Krippendorff’s $\alpha$ on pre-adjudication annotations. As shown in Table~\ref{tab:annotation_iaa}, agreement varies across dimensions: perceptual speech quality and audio-visual lip-sync achieve the highest agreement, while more subjective dimensions such as affective and expressive alignment show comparatively lower consistency. Overall, the annotation results indicate reliable labeling quality across the benchmark.

\begin{table}[t]
\centering
\small
\setlength{\tabcolsep}{8pt}
\renewcommand{\arraystretch}{1.05}
\caption{Pre-adjudication inter-annotator agreement measured by Krippendorff's $\alpha$.}
\label{tab:annotation_iaa}
\begin{tabular}{lc}
\toprule
\textbf{Dimension} & \textbf{$\alpha$} \\
\midrule
VQ & 0.76 \\
SQ & 0.88 \\
SC & 0.78 \\
CC & 0.79 \\
LS & 0.82 \\
SA & 0.68 \\
TT & 0.71 \\
EA & 0.65 \\
CA & 0.80 \\
\bottomrule
\end{tabular}
\end{table}

\section{Description for Failure Modes}
\label{subsec:failure_modes}

Tables~\ref{tab:failure_quality}--\ref{tab:failure_consistency} present the fine-grained failure taxonomy in our benchmark. They cover quality-level defects (Table~\ref{tab:failure_quality}), cinematic-level failures (Table~\ref{tab:failure_cinematic}), interaction-level errors (Table~\ref{tab:failure_interaction}), and consistency-level failures (Table~\ref{tab:failure_consistency}). Together, these tables define the evaluation scope and provide the basis for structured annotation and diagnosis.

\section{Visualizations of failure modes}
\label{subsec:failure_visualizations}

\begin{figure*}
    \centering
    \includegraphics[width=0.9\textwidth]{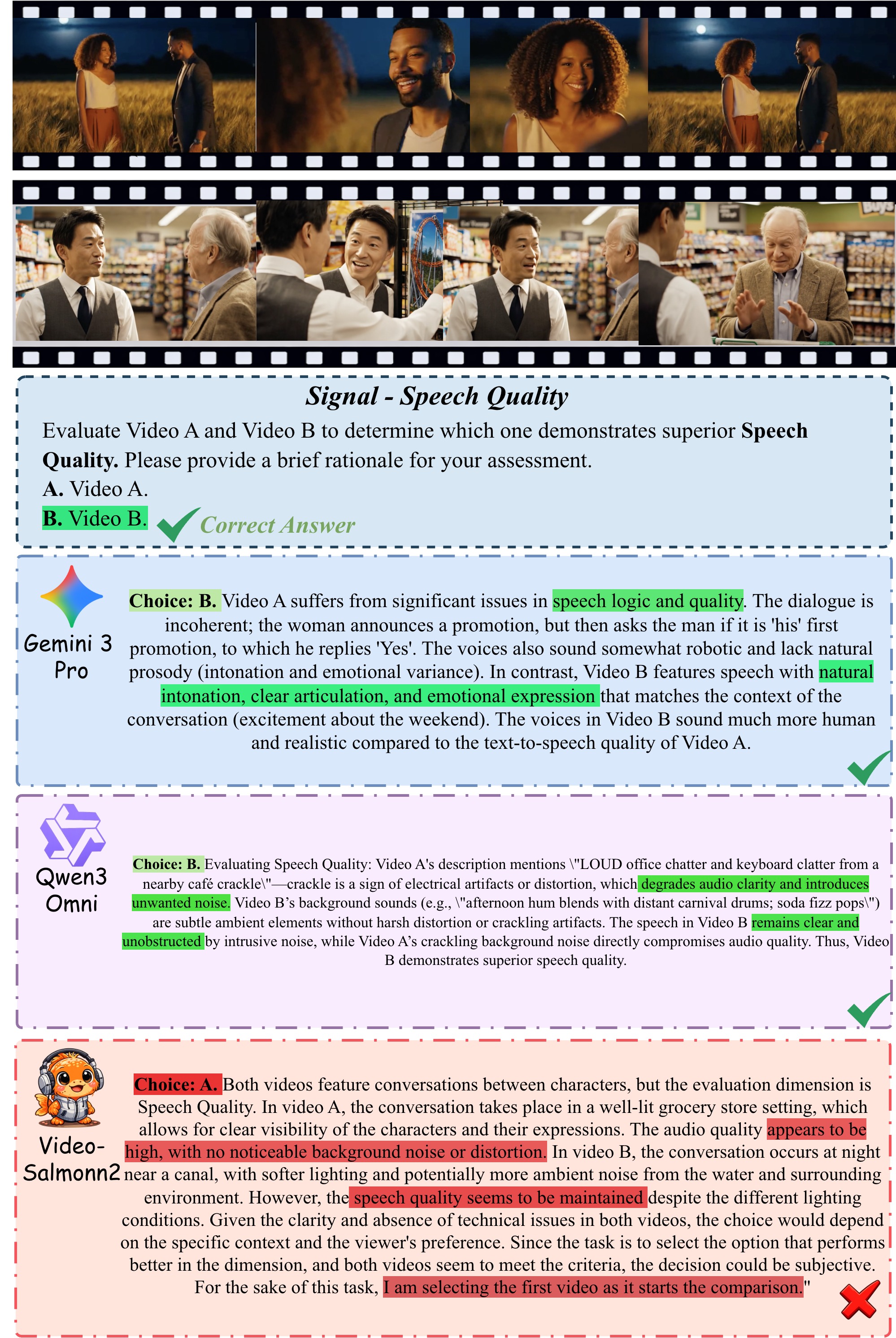}
    \caption{Case Study - Signal Level}
    \label{fig:case-sq}
\end{figure*}

\begin{figure*}
    \centering
    \includegraphics[width=0.9\textwidth]{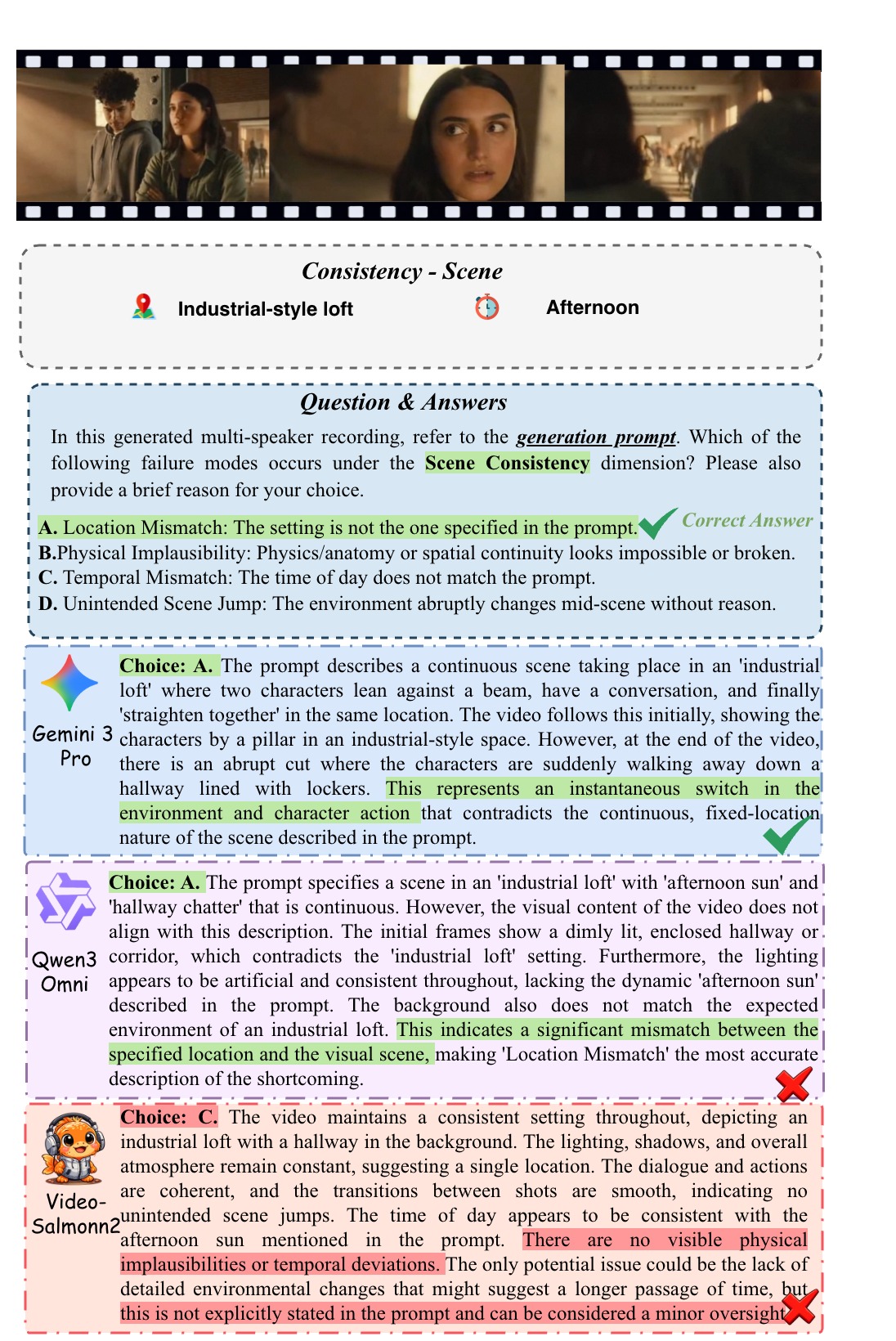}
    \caption{Case Study - Consistency Level}
    \label{fig:case-consist}
\end{figure*}

\begin{figure*}
    \centering
    \includegraphics[width=0.9\textwidth]{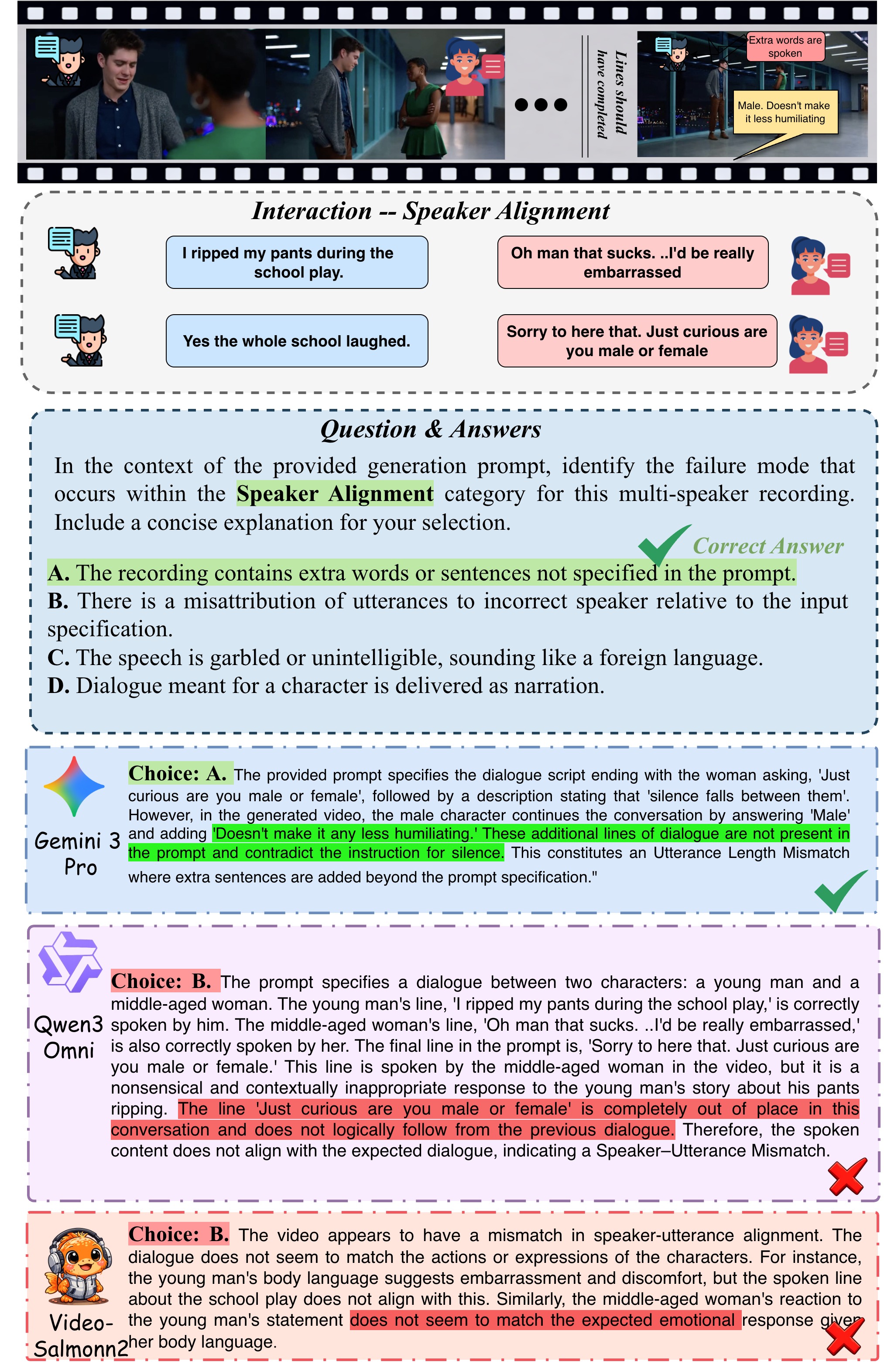}
    \caption{Case Study - Interaction Level}
    \label{fig:case-3}
\end{figure*}

\begin{figure*}
    \centering
    \includegraphics[width=0.9\textwidth]{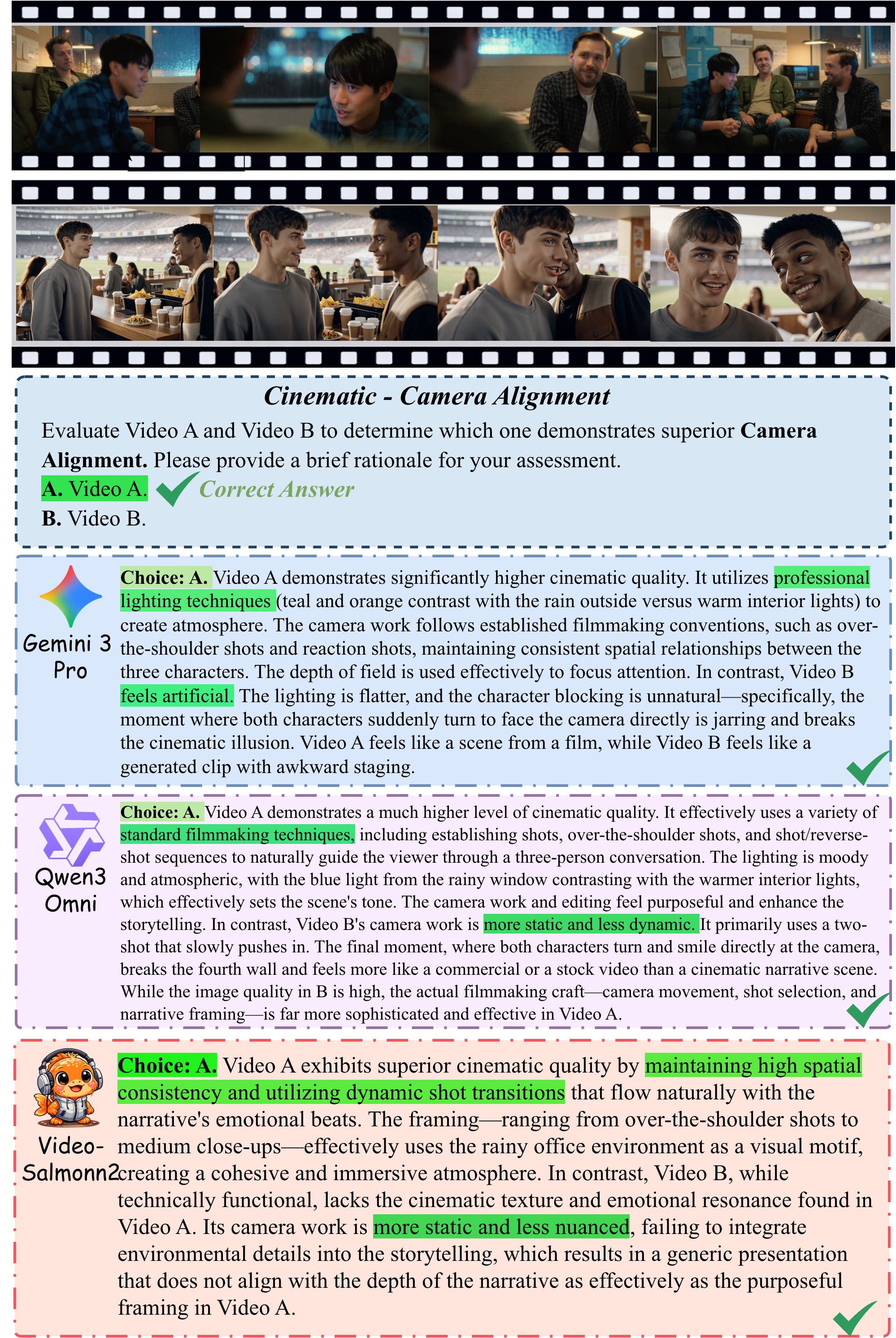}
    \caption{Case Study - Cinematic Level}
    \label{fig:case-4}
\end{figure*}

\begin{figure*}[t]
  \centering
  \includegraphics[width=\textwidth]{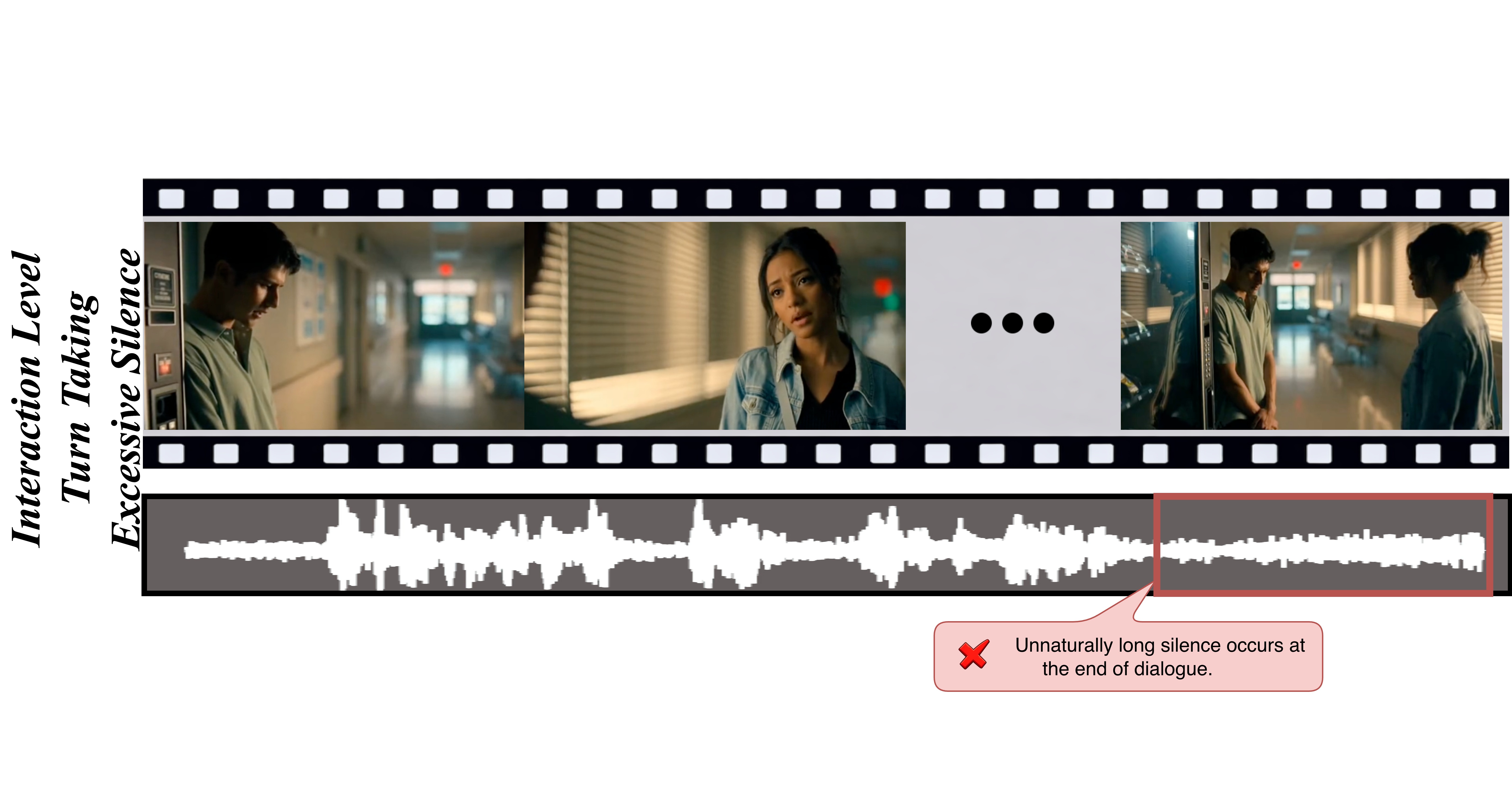}
  \caption{Visualizations of failure mode - Excessive Silence}
  \label{fig:failure_mode_excessive_silence}
\end{figure*}

\begin{figure*}[t]
  \centering
  \includegraphics[width=\textwidth]{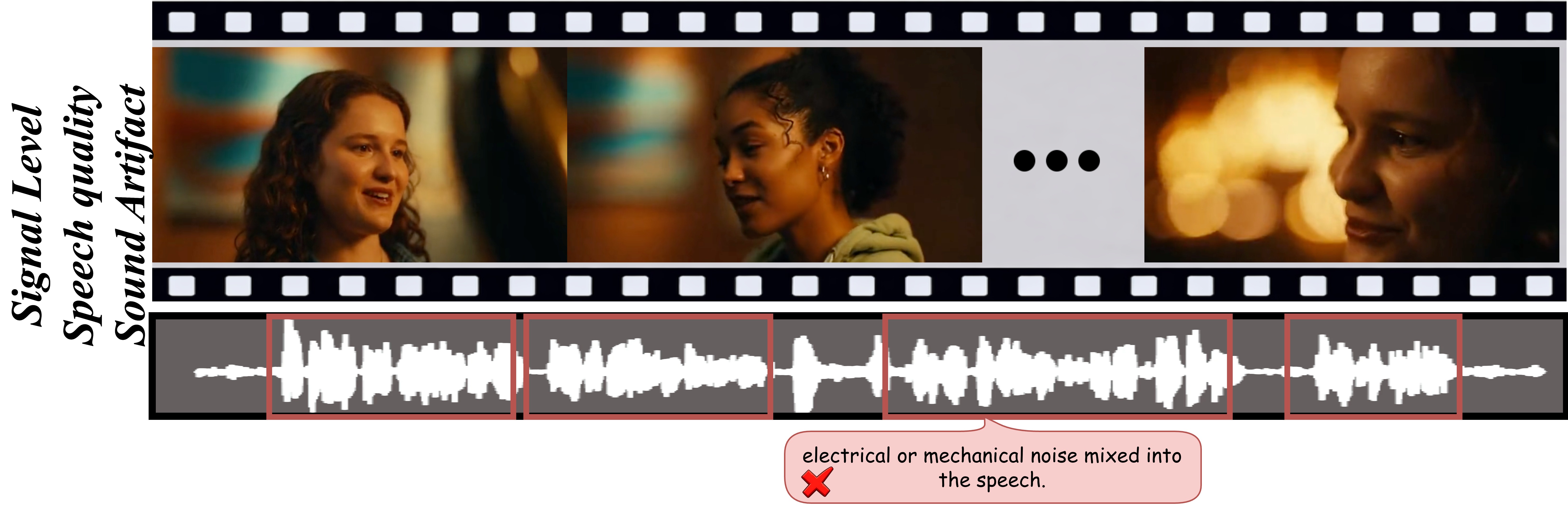}
  \caption{Visualizations of failure mode - Sound Artifact}
  \label{fig:failure_mode_sound_artifact}
\end{figure*}

\begin{figure*}[t]
  \centering
  \includegraphics[width=\textwidth]{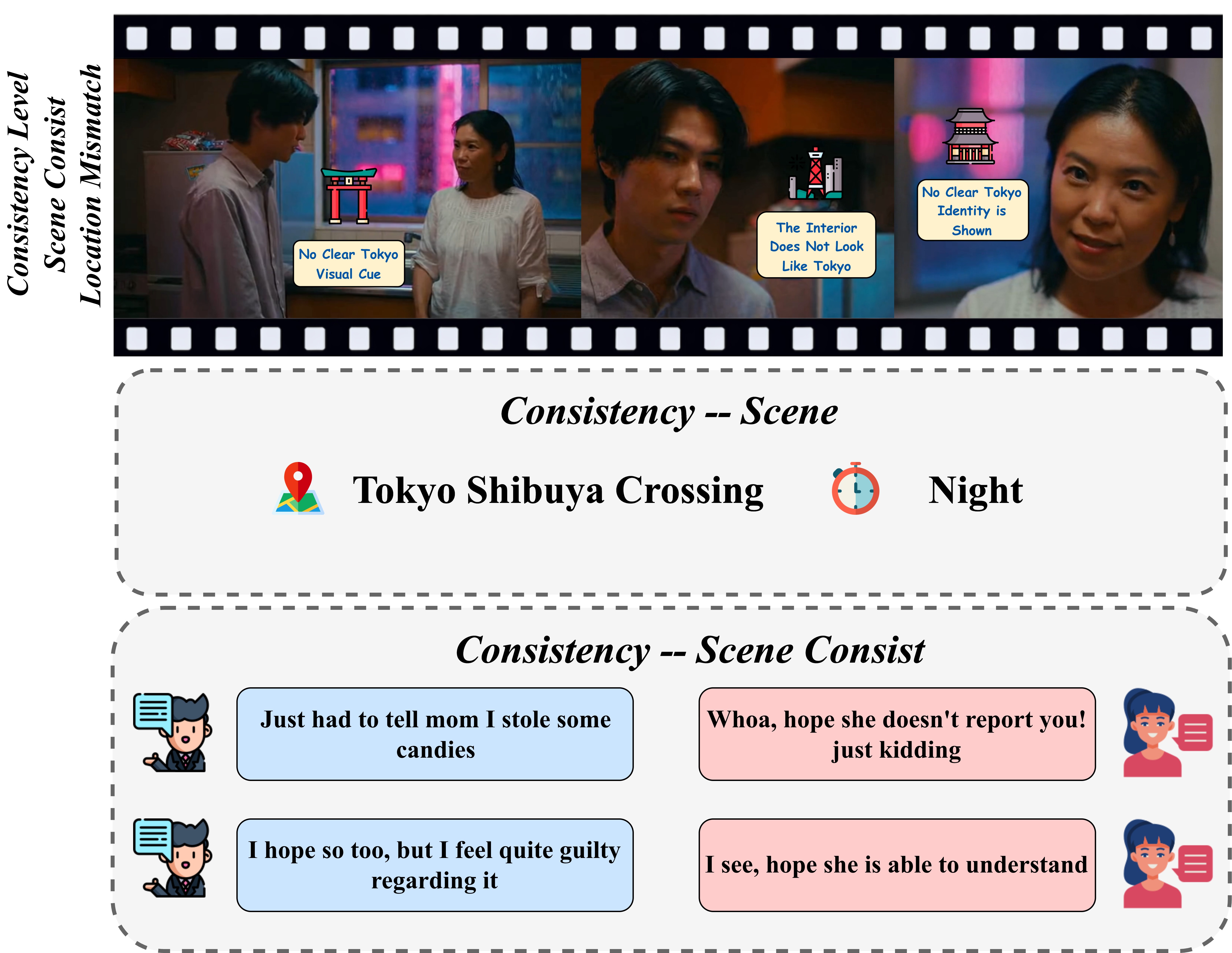}
  \caption{Visualizations of failure mode - Location Mismatch}
  \label{fig:failure_mode_location_mismatch}
\end{figure*}

\begin{figure*}[t]
  \centering
  \includegraphics[width=\textwidth]{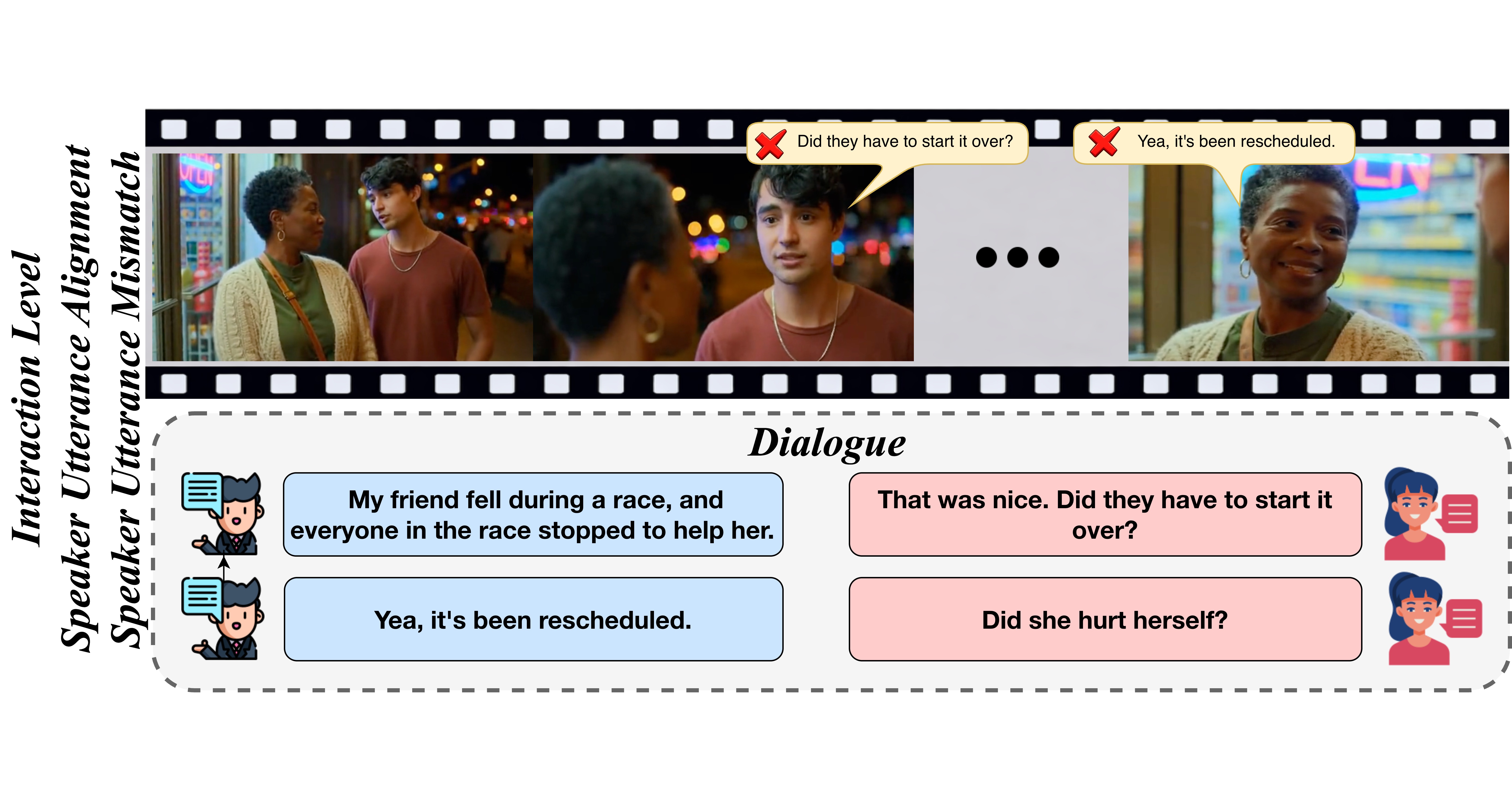}
  \caption{Visualizations of failure mode - Speaker Utterance Mismatch}
  \label{fig:failure_mode_spk_utt_mismatch}
\end{figure*}

\begin{figure*}[t]
  \centering
  \includegraphics[width=\textwidth]{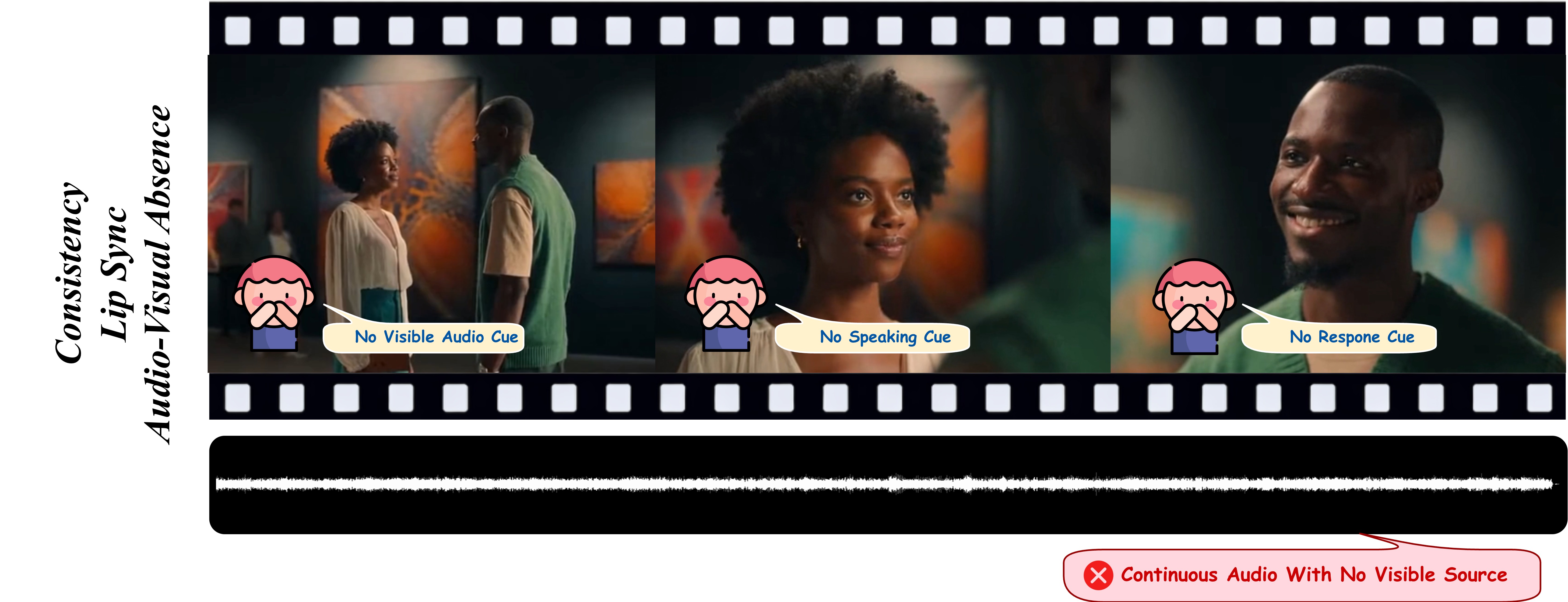}
  \caption{Visualizations of failure mode - Audio-Visual Absence}
  \label{fig:failure_mode_audio_visual_absence}
\end{figure*}

\begin{figure*}[t]
  \centering
  \includegraphics[width=\textwidth]{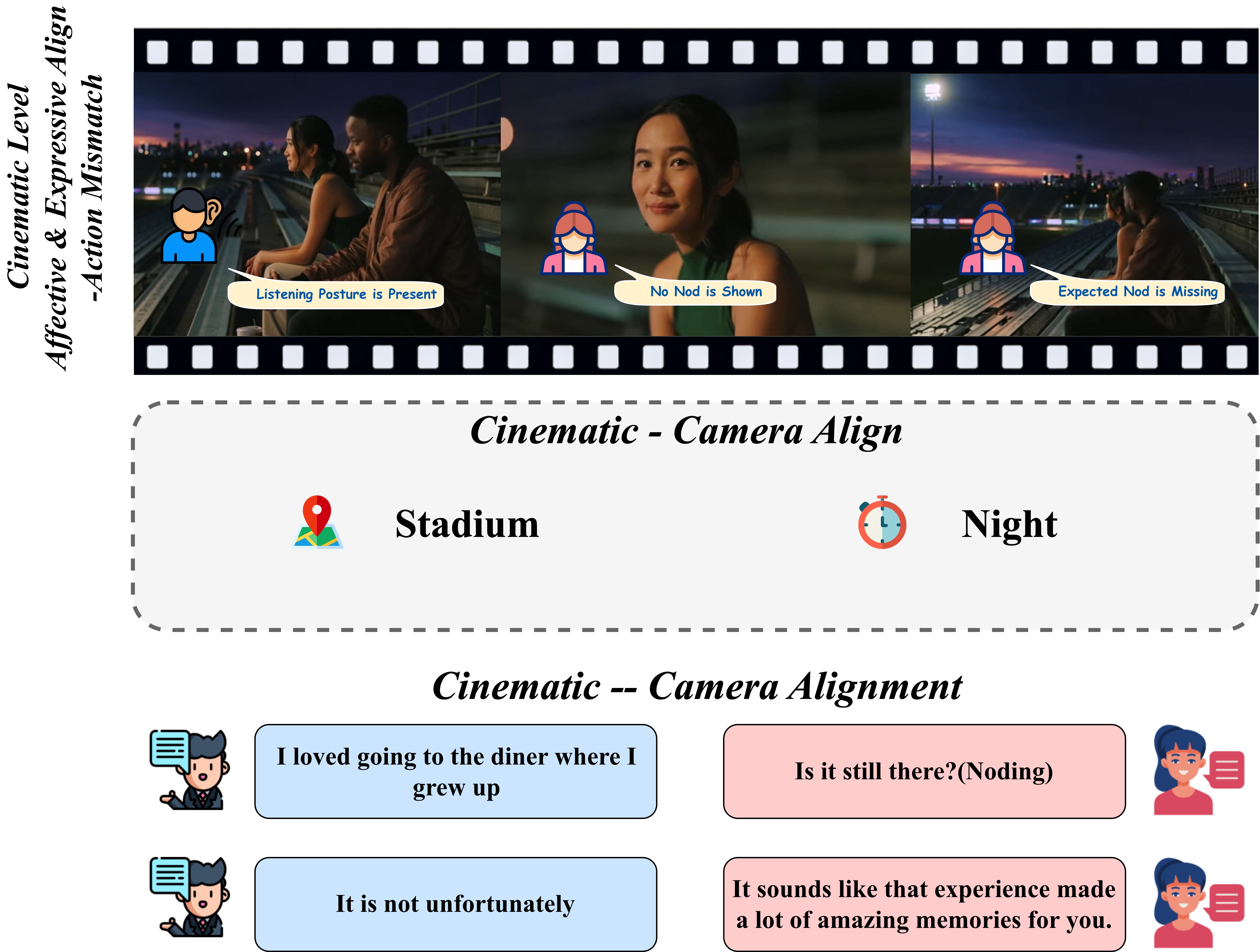}
  \caption{Visualizations of failure mode - Action Mismatch}
  \label{fig:failure_mode_action_mismatch}
\end{figure*}

\begin{figure*}[t]
  \centering
  \includegraphics[width=\textwidth]{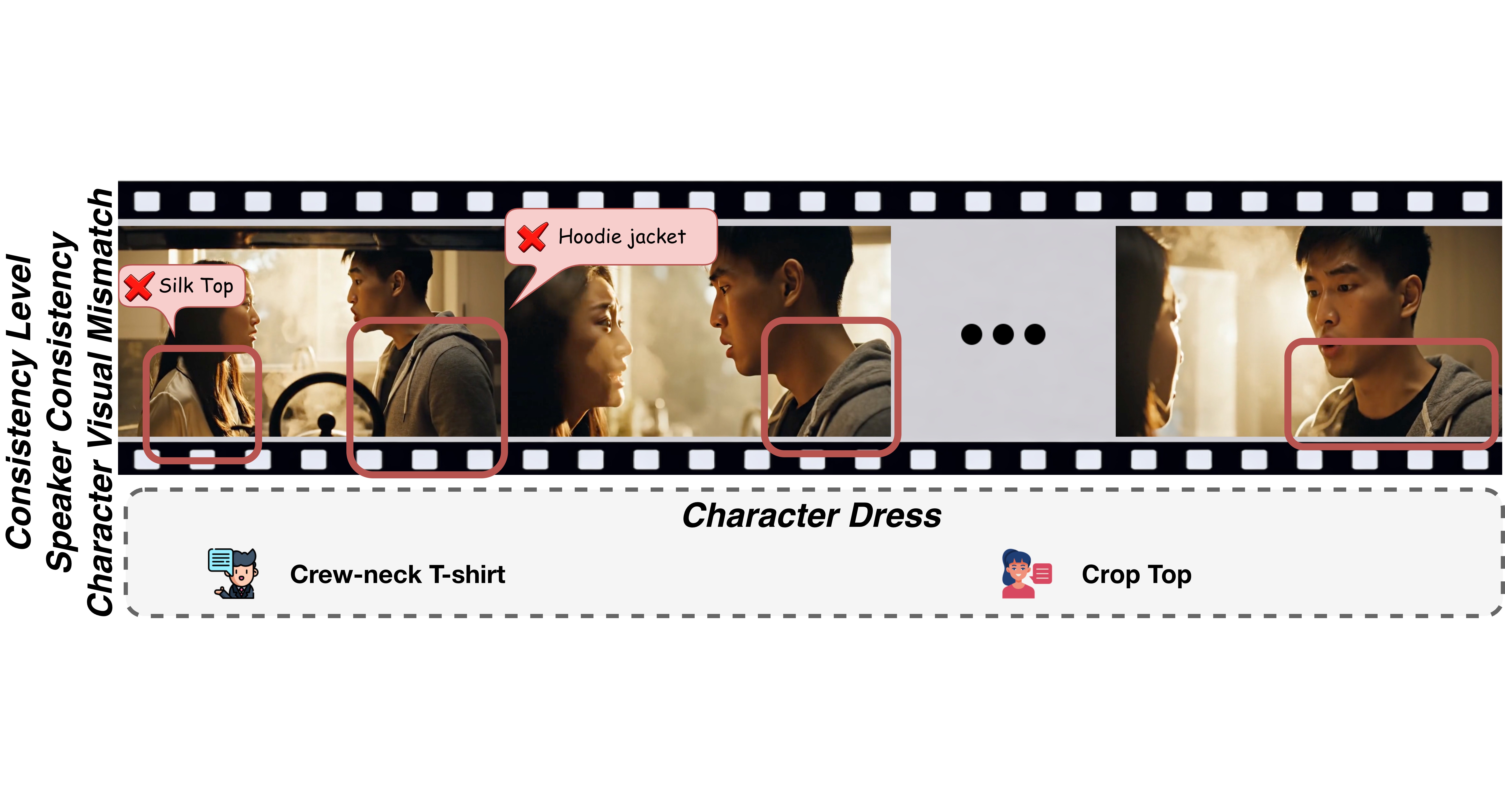}
  \caption{Visualizations of failure mode - Character Visual Mismatch}
  \label{fig:failure_mode_char_visual_mismatch}
\end{figure*}

\begin{figure*}[t]
  \centering
  \includegraphics[width=\textwidth]{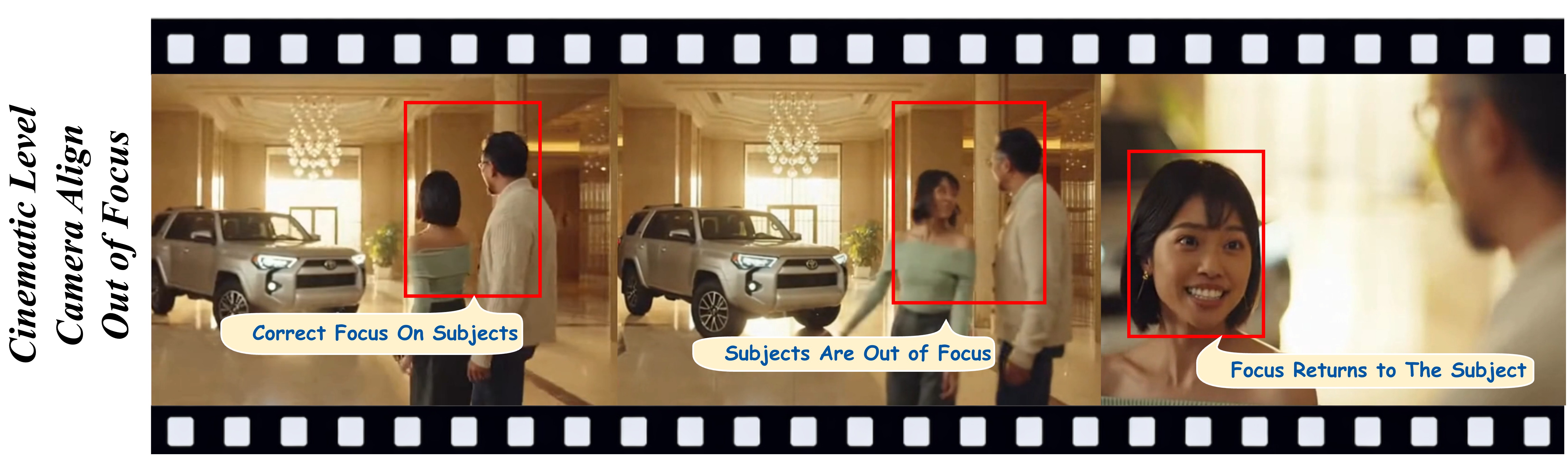}
  \caption{Visualizations of failure mode - Out of Focus}
  \label{fig:failure_mode_out_of_focus}
\end{figure*}

Figures~\ref{fig:failure_mode_excessive_silence}--\ref{fig:failure_mode_out_of_focus} present representative examples of fine-grained failure modes across four levels. In Figures~\ref{fig:case-sq}--\ref{fig:case-4}, each case is accompanied by the MLLM's choice and rationale, providing qualitative evidence of how different failure patterns are identified and interpreted in practice.

\section{Overall Quality Assessment} \label{subsec:quality_overall}

Beyond fine-grained failure-mode diagnosis, we also assess overall quality under the natural generation distribution using an automatic evaluation toolkit, embedding-based methods (e.g., ImageBind~\cite{girdhar2023imagebind}), and MLLM-as-a-judge. Specifically, we conduct this holistic evaluation on unfiltered samples generated by Sora~\cite{openai2025sora2}, Veo~\cite{google2025veo3}, Wan~\cite{alibaba2025wan25}, and the open-source models Ltx2.3~\cite{hacohen2026ltx} and Ovi~\cite{low2025ovi}. For each sample, we instantiate the MLLM judge with Gemini 3 Pro, which was identified as the strongest model in our earlier failure diagnosis experiments, and use the dimension-specific evaluation prompts shown in Figures~\ref{fig:prompt_design_scene_consistency}--\ref{fig:prompt_design_camera_composition_alignment} to score scene consistency, character consistency, speaker--utterance alignment, turn-taking logic, affective and expressive alignment, and camera/composition alignment.

This protocol complements failure diagnosis by allowing us to more holistically examine the preferences of generation models under the natural generation distribution. However, such methods are still limited in their ability to provide sufficiently fine-grained analysis. This highlights the need for dedicated diagnostic models that can support more fine-grained evaluation of audiovisual video quality in the future. As shown in Table~\ref{tab:natural_dist_eval}, current generators generally achieve relatively high scores on perceptual and low-level quality dimensions, but still lag behind on dialogue- and interaction-centric dimensions.

For metrics such as scene consistency, we report not only MLLM-based scores but also objective measures. Speech quality is evaluated using the AudioBox Aesthetics model. Cross-modal alignment scores are computed by extracting modality representations with ImageBind and measuring their cosine similarity, including text--video (T--V), text--audio (T--A), and audio--visual (A--V) alignment. These metrics provide complementary signals for assessing multimodal consistency.

It is worth noting that MLLM-as-a-judge tends to assign relatively high overall scores. Therefore, these results are better interpreted as relative comparisons across models and dimensions, rather than as absolute estimates of quality or of the true state of audiovisual video generation. Nevertheless, weaknesses in interaction-related and higher-level reasoning dimensions remain persistent. This suggests that the structural failure patterns revealed by MTAVG-Bench also recur under the natural generation distribution, further underscoring the need for a stronger Omni diagnostic model with more fine-grained recognition capabilities for audiovisual generation.

\newpage

\begin{table*}[t]
\centering
\small
\setlength{\tabcolsep}{4pt}
\renewcommand{\arraystretch}{1.12}
\caption{Failure modes under the \textbf{Quality Level}.}
\label{tab:failure_quality}
\begin{tabular}{%
  p{0.18\textwidth}
  p{0.24\textwidth}
  p{0.50\textwidth}
}
\toprule
\textbf{Sub-dimension} & \textbf{Failure Mode} & \textbf{Description} \\
\midrule
\multirow{3}{=}{\textbf{Speech Quality}}
& \textbf{Environmental Noise}
& Audible background noise is present in the generated audio and interferes with the listener's ability to clearly perceive or understand the speech signal. \\
\cmidrule(lr){2-3}
& \textbf{Sound Artifact}
& Unnatural non-speech sounds, such as electrical noise, mechanical noise, buzzing, distortion, or other synthetic artifacts, are mixed into the speech audio. \\
\cmidrule(lr){2-3}
& \textbf{Audio Interruption}
& The speech signal contains abnormally long silences, stuttering, or other interruptions that disrupt the temporal continuity and fluency of the spoken utterance. \\
\midrule

\multirow{3}{=}{\textbf{Video Quality}}
& \textbf{Visual Degradation}
& The generated video exhibits perceptible degradation in visual quality, including blur, noise, compression-like artifacts, or loss of fine details that reduces visual clarity. \\
\cmidrule(lr){2-3}
& \textbf{Temporal Flickering}
& The video lacks temporal coherence across consecutive frames, causing flickering, jittering, or unstable appearance in speaker regions, object boundaries, or background areas. \\
\cmidrule(lr){2-3}
& \textbf{Geometric Clipping}
& Incorrect geometric rendering causes characters or objects to be clipped, intersect improperly, or appear with missing parts, resulting in visibly invalid spatial structure. \\
\bottomrule
\end{tabular}
\end{table*}

\begin{table*}[t]
\centering
\small
\setlength{\tabcolsep}{4pt}
\renewcommand{\arraystretch}{1.12}
\caption{Failure modes under the \textbf{Consistency Level}.}
\label{tab:failure_consistency}
\begin{tabular}{%
  p{0.18\textwidth}
  p{0.24\textwidth}
  p{0.50\textwidth}
}
\toprule
\textbf{Sub-dimension} & \textbf{Failure Mode} & \textbf{Description} \\
\midrule
\multirow{5}{=}{\textbf{Lip Sync}}
& \textbf{Lip-Sync Mismatch}
& The visible lip articulation of a speaker is not temporally aligned with the corresponding speech signal, such that the mouth movements occur noticeably earlier or later than the associated audio. \\
\cmidrule(lr){2-3}
& \textbf{Missing Lip Motion}
& The speech signal is present and attributable to a visible speaker, but the speaker's lips remain static or fail to exhibit the corresponding articulatory movement. \\
\cmidrule(lr){2-3}
& \textbf{Phantom Lip Motion}
& The speaker's lips exhibit visible articulatory movement even though no corresponding speech signal is present in the audio. \\
\cmidrule(lr){2-3}
& \textbf{Multi-Lip Single Voice}
& A single speech signal is incorrectly synchronized with the lip movements of multiple visible speakers, creating the appearance that more than one mouth is producing the same voice. \\
\cmidrule(lr){2-3}
& \textbf{Audio-Visual Absence}
& A visible speaker is expected to speak, but both the speech audio and the corresponding lip movement are absent, leaving the speaker without either auditory or visual speech cues. \\
\midrule

\multirow{5}{=}{\textbf{Character Consistency}}
& \textbf{Character Visual Mismatch}
& At least one speaker is rendered with visual character attributes that are inconsistent with the prompt specification, including but not limited to gender, age, race, clothing, or other explicitly described appearance cues. \\
\cmidrule(lr){2-3}
& \textbf{Character Anomaly}
& The character exhibits implausible visual appearance, behavior, or temporal continuity, including visual artifacts, unnatural motion, sudden disappearance or reappearance, abrupt changes in presence, or transformations that are not explained by the prompt. \\
\cmidrule(lr){2-3}
& \textbf{Speaker--Voice Attribute Mismatch}
& The speaker's vocal timbre is inconsistent with the demographic or character attributes visually represented on screen, causing the perceived voice to conflict with the character's apparent identity. \\
\cmidrule(lr){2-3}
& \textbf{Speaker Voice Inconsistency}
& The speaker's voice characteristics are not stable over time or contain localized segment-level anomalies, resulting in inconsistent vocal identity or abnormal voice quality within the same speaker. \\
\cmidrule(lr){2-3}
& \textbf{Character Visual Artifact}
& The character displays implausible visual appearance or behavior, such as distorted facial features, unnatural body motion, malformed anatomy, or visible rendering artifacts affecting the character. \\
\midrule

\multirow{4}{=}{\textbf{Scene Consistency}}
& \textbf{Location Mismatch}
& The depicted environment does not correspond to the location specified in the prompt, resulting in a scene setting that conflicts with the intended spatial context. \\
\cmidrule(lr){2-3}
& \textbf{Physical Implausibility}
& The scene violates basic physical or commonsense constraints, including spatial, temporal, or causal inconsistencies that make the depicted event implausible. \\
\cmidrule(lr){2-3}
& \textbf{Unintended Scene Jump}
& The entire environment changes instantaneously even though the dialogue or action is intended to proceed continuously within a fixed location, producing an unintended discontinuity in scene setting. \\
\cmidrule(lr){2-3}
& \textbf{Temporal Mismatch}
& The depicted time or temporal setting is inconsistent with the prompt specification, such as when the generated scene reflects an unintended time of day, period, or temporal context. \\
\bottomrule
\end{tabular}
\end{table*}

\begin{table*}[t]
\centering
\small
\setlength{\tabcolsep}{4pt}
\renewcommand{\arraystretch}{1.12}
\caption{Failure modes under the \textbf{Interaction Level}.}
\label{tab:failure_interaction}
\begin{tabular}{%
  p{0.18\textwidth}
  p{0.24\textwidth}
  p{0.50\textwidth}
}
\toprule
\textbf{Sub-dimension} & \textbf{Failure Mode} & \textbf{Description} \\
\midrule
\multirow{6}{=}{\textbf{Turn-Taking}}
& \textbf{Speaker--Utterance Mismatch}
& Spoken lines are incorrectly attributed to characters, causing an utterance intended for one speaker to be delivered by another speaker. \\
\cmidrule(lr){2-3}
& \textbf{Utterance Length Mismatch (Missing)}
& Required words or sentences specified in the prompt are omitted from the generated speech, resulting in an incomplete realization of the intended utterance. \\
\cmidrule(lr){2-3}
& \textbf{Utterance Length Mismatch (Extra)}
& Additional words or sentences that are not specified in the prompt are inserted into the generated speech, resulting in content that exceeds the intended utterance. \\
\cmidrule(lr){2-3}
& \textbf{Narration Shift}
& Dialogue intended to be spoken directly by a character is instead delivered as narration, changing the intended mode of speech presentation. \\
\cmidrule(lr){2-3}
& \textbf{Semantic Deviation}
& The spoken content substantially deviates from the intended semantic meaning of the prompt, resulting in speech that conveys a different or incorrect meaning. \\
\cmidrule(lr){2-3}
& \textbf{Pronunciation Error}
& The speech contains unintelligible or corrupted phonetic patterns that resemble a foreign language, making the intended utterance difficult or impossible to understand. \\
\midrule

\multirow{4}{=}{\textbf{Speaker--Utterance Alignment}}
& \textbf{Utterance Length Mismatch}
& The speech is abruptly truncated in the middle of a sentence or utterance, preventing the intended utterance from being completed and disrupting the continuity of the dialogue. \\
\cmidrule(lr){2-3}
& \textbf{Speaker Turn Swap and Failure}
& Dialogue coherence breaks down because the model confuses speaker turns, produces abrupt or unnatural transitions between turns, or introduces extraneous participants not intended by the prompt. \\
\cmidrule(lr){2-3}
& \textbf{Overlapping Speech}
& Interruptions or simultaneous speaking occur between speakers, causing vocal interference and reducing the clarity of the dialogue. \\
\cmidrule(lr){2-3}
& \textbf{Excessive Silence}
& Unnaturally long silence occurs at the beginning or end of the dialogue, creating an abnormal pause that disrupts the expected conversational timing. \\
\bottomrule
\end{tabular}
\end{table*}

% Failure mode Description
\begin{table*}[t]
\centering
\small
\setlength{\tabcolsep}{4pt}
\renewcommand{\arraystretch}{1.12}
\caption{Failure modes under the \textbf{Cinematic Level}.}
\label{tab:failure_cinematic}
\begin{tabular}{%
  p{0.18\textwidth}
  p{0.24\textwidth}
  p{0.50\textwidth}
}
\toprule
\textbf{Sub-dimension} & \textbf{Failure Mode} & \textbf{Description} \\
\midrule
\multirow{4}{=}{\textbf{Affect Expression}}
& \textbf{Action Mismatch}
& The body actions displayed by the character are inconsistent with the actions specified in the prompt, causing the generated behavior to contradict the intended action description. \\
\cmidrule(lr){2-3}
& \textbf{Behavioral Rigidity}
& Speech delivery lacks sufficient prosodic variation or emotional expressiveness, or the character's movement and demeanor appear unnaturally rigid, mechanical, or atypical, resulting in behavior that lacks natural expressiveness. \\
\cmidrule(lr){2-3}
& \textbf{Emotion Mismatch}
& The emotional expression displayed by the character is excessively exaggerated, distorted, or inconsistent with the prompt, resulting in an affective state that does not match the intended emotion. \\
\cmidrule(lr){2-3}
& \textbf{Spatial Inconsistency}
& The speaker provides insufficient expressive responses to the conversational context, such as missing or weak turn-taking cues, feedback gestures, or engagement signals during interaction. \\
\midrule

\multirow{2}{=}{\textbf{Camera Alignment}}
& \textbf{Out of Focus}
& The camera focus is inadequately controlled, causing the intended subject to appear blurred or out of focus when it should be visually clear. \\
\cmidrule(lr){2-3}
& \textbf{Camera Framing Failure}
& The camera framing or motion is improperly aligned with the active speaker, causing the speaker to be poorly positioned, partially excluded, or visually miscentered in the shot. \\
\bottomrule
\end{tabular}
\end{table*}

\clearpage
\begin{figure*}[]
    \centering
    
    \begin{systemprompt}
        
        (Audio-Video Generation Prompt)\\
        
        \textit{You are a professional video generation script designer}. Your task is to transform specific input data into a high-fidelity, realistic cinematic narrative prompt. You create vivid, lifelike scenes that capture the exact emotional essence of the provided context while adhering to strict technical and character requirements.\\
        
        \#\#\# 1. Input Variable Handling.\\
        You will receive data in the following format:
        \begin{itemize}[leftmargin=*, topsep=2pt, itemsep=0pt]
            \item \textbf{context}: [The original emotional state, mood, or setting].
            \item \textbf{conversation}: [The raw dialogue exchange between characters].
        \end{itemize}
        
        \#\#\# 2. Character \& Environmental Specifications. You must define the following with high specificity in every prompt:\\
        - \textbf{Character Attributes}: For every character, you must specify their \textbf{gender, age} (e.g., mid-20s, elderly), \textbf{race/ethnicity}, and \textbf{detailed dress/clothing} (e.g., "a faded denim jacket over a white tee," "a sharp pinstripe charcoal suit").\\
        - \textbf{Environment}: Define a specific \textbf{location} and the \textbf{time of day} (Morning, Afternoon, Evening, or Night).\\
        - \textbf{Visual Style}: The style must always be \textbf{Realistic} or \textbf{Hyper-Realistic}, emphasizing natural textures, cinematic lighting, and authentic skin details.\\
        - \textbf{Ambiance}: Describe a dominant sound and lighting condition that directly supports the \textbf{context} (e.g., "The distant drone of a city" for a lonely context, or "Warm amber glow" for a nostalgic context).\\
        
        \#\#\# 3. The Speaker-Centric Camera Rule (Mandatory)\\
        The camera must focus on the person currently speaking. You must integrate cinematic tags inside square brackets \texttt{[...]} at the exact moment the dialogue shifts:\\
        - \textbf{Speaker Focus \& Cinematic Variety}: Use \texttt{[Focus on One]}, \texttt{[Close shot]}, \texttt{[Medium close shot]}, \texttt{[Master Shot]}, \texttt{[Two Shot]} or \texttt{[Shot/Reverse Shot]} while ensuring the active speaker is the visual centerpiece.\\
        
        \#\#\# 4. Narrative \& Dialogue Logic\\
        - \textbf{Maintain Original Meaning}: Do \textbf{not} rewrite the mood to be "positive" unless the context is already positive. Keep the emotional arc and semantics exactly as provided in the \texttt{context} and \texttt{conversation}.\\
        - \textbf{Dialogue Format}: Use the dialogue from the input exactly. Enclose all spoken lines in \textbf{single quotes ('...')}.\\
        - \textbf{Visual Action}: Include non-verbal cues (gestures, expressions) that match the emotional \texttt{context}.\\
        
        \#\#\# 5. Final Output Constraints\\
        - \textbf{Direct Output Only}: Output \textbf{only} the final, continuous natural language prompt. No JSON, no labels, no headers, and no introductory filler.\\
        - \textbf{Format}: A fluid, evocative narrative that is easy to copy and use for video generation.\\
        
        ---\textbf{TASK:} Receive the \texttt{context} and \texttt{conversation} and generate the professional cinematic narrative prompt now.
    \end{systemprompt}
    \caption{The specific system prompt for decomposing text descriptions into hierarchical semantic levels.}
    \label{fig:prompt_design_generating}
\end{figure*}

\begin{figure*}[h]
    \centering
    
    \begin{systemprompt}
        
        (Single Choice Question Inference)\\
        
        \#\# Role:\\
        \textit{You are a Senior Diagnostic Auditor for AIGC-generated video}, specializing in the forensic analysis of multi-talker dialogue generation. Your expertise lies in performing side-by-side diagnostic comparisons of multi-talker dialogue videos to determine which one better aligns with real-world physical and social dynamics.\\
        
        \#\# Evaluation Context:\\
        1. \textbf{Target Evaluation Dimension:} \{dimension\_definition\}\\
        2. \textbf{Generation Intent (Prompt):} \{video\_prompt\}\\
        3. \textbf{Evidence (Video):} \{video\_path\}\\
        
        \#\# Task\\
        Based on the Evaluation Dimension provided, analyze the video and answer the following diagnostic question:\\
        \textbf{Question:} \{question\}\\
        
        \#\# Operational Requirements:\\
        1. \textbf{Dimension Constraints:} Evaluate the video strictly through the lens of the provided \{dimension\_definition\}. Ignore issues unrelated to this specific metric.\\
        2. \textbf{Forensic Evidence:} In your reasoning, cite specific visual or auditory evidence (e.g., \textit{"Speaker A's mouth remains closed while their voice is heard,"} or "The camera fails to switch to the active speaker").\\
        
        \#\# Output Format\\
        You must return a \textbf{strictly formatted JSON object}. No markdown code blocks, no conversational filler, only the raw JSON.\\
        
        \texttt{\{\\
        \hspace*{1em} "reasoning": "Reasoning for each option step by step, citing specific forensic evidence.",\\
        \hspace*{1em} "choice": "The letter of the selected option (e.g., 'A')"\\
        \}}
    \end{systemprompt}
    \caption{The specific system prompt for Single Choice Question Inference.}
    \label{fig:prompt_design_inference}
\end{figure*}

\begin{figure*}[h]
    \centering
    % Figure 3: Diagnostic Specialist (Alternative Version)
    \begin{systemprompt}
        \# Role:\\
        Diagnostic Specialist for AIGC-generated Video\\
        
        \#\# Description:\\
        Your task is to identify specific failures in AI-generated dialogue videos by analyzing the alignment between user intent and multi-modal output.\\
        
        \#\# Evaluation Context\\
        1. Target Evaluation Dimension: \{dimension\_definition\}\\
        2. Generation Intent (Prompt): \{video\_prompt\}\\
        3. Evidence (Video): \{video\_path\}\\
          
        \#\# Task\\
        Based on the Evaluation Dimension provided, analyze the video and answer the following diagnostic question:\\
        Question: \{question\}\\
        
        \#\# Operational Requirements\\
        1. Dimension Constraints: Evaluate the video strictly through the lens of the provided \texttt{\{dimension\_definition\}}. Ignore issues unrelated to this specific metric.\\
        2. Forensic Evidence: In your reasoning, cite specific visual or auditory evidence (e.g., "Speaker A's mouth remains closed while their voice is heard," or "The camera fails to switch to the active speaker").\\
          
        \#\# Output Format\\
        You must return a strictly formatted JSON object. No markdown code blocks, no conversational filler, only the raw JSON.\\
        
        \texttt{\{\\
        \hspace*{1em} "reasoning": "Reasoning for each option step by step, citing specific forensic evidence.",\\
        \hspace*{1em} "choice": "The letter of the selected option (e.g., 'A')"\\
        \}}
    \end{systemprompt}
    \caption{The specific system prompt for Diagnostic Specialist analysis.}
    \label{fig:prompt_design_specialist}
\end{figure*}

\begin{figure*}[t]
    \centering
    \setlength{\textfloatsep}{6pt}
    \begin{systemprompt}
    \small
    (Audio-Visual Evaluation Prompt: Scene Consistency)\\
    \textit{You are a professional audio-visual scene coherence evaluator}. Your task is to evaluate whether an AI-generated 8--15 second audiovisual video presents a coherent, plausible, and temporally stable scene that matches the provided generation prompt. Focus only on \textbf{Scene Consistency (SC)}.\\

    \#\#\# 1. Input Variable Handling.\\
    You will receive data in the following format:
    \begin{itemize}[leftmargin=*, topsep=1pt, itemsep=0pt, parsep=0pt, partopsep=0pt]
        \item \textbf{Dialogue Context}: [Overall emotional state, mood, or situational context].
        \item \textbf{Dialogue Scene}:
        \begin{itemize}[leftmargin=1.5em, topsep=1pt, itemsep=0pt, parsep=0pt, partopsep=0pt]
            \item \textbf{Location}: [Where the scene takes place].
            \item \textbf{Time}: [Time of day or temporal setting].
            \item \textbf{Description}: [Scene-level environmental description].
        \end{itemize}
        \item \textbf{Characters}: [Character specifications, provided for reference only when needed for scene interpretation].
        \item \textbf{Dialogue Content}: [The dialogue spoken in the video].
        \item \textbf{Video}: [An 8--15 second generated audio-video].
    \end{itemize}

    \#\#\# 2. Evaluation Focus.\\
    Evaluate whether the \textbf{Video} faithfully and consistently realizes the scene specified by \textbf{Dialogue Context} and \textbf{Dialogue Scene}:\\
    \begin{itemize}[leftmargin=*, topsep=1pt, itemsep=0pt, parsep=0pt, partopsep=0pt]
        \item \textbf{Location Consistency}: Does the visible and audible environment match the specified \textbf{Location}?
        \item \textbf{Time Consistency}: Do lighting, atmosphere, and environmental cues match the specified \textbf{Time}?
        \item \textbf{Description Realization}: Does the scene in the \textbf{Video} reflect the environmental details described in \textbf{Dialogue Scene.Description}?
        \item \textbf{Scene Plausibility}: Are background, props, ambient sound, and spatial layout mutually compatible?
        \item \textbf{Temporal Scene Stability}: Does the scene remain logically stable over the full clip without abrupt unexplained shifts in environment, place, or atmosphere?
        \item \textbf{Contextual Fit}: Does the overall scene support the situation implied by \textbf{Dialogue Context} and \textbf{Dialogue Content}?
    \end{itemize}

    \#\#\# 3. Scoring Standard (1--5).\\
    Assign an integer score from 1 to 5:
    \begin{itemize}[leftmargin=*, topsep=1pt, itemsep=0pt, parsep=0pt, partopsep=0pt]
        \item \textbf{5}: The scene is fully coherent, plausible, and consistently aligned with \textbf{Dialogue Context} and \textbf{Dialogue Scene}.
        \item \textbf{4}: The scene is mostly coherent, with only minor inconsistencies that do not substantially harm realism or contextual fit.
        \item \textbf{3}: The scene is understandable but contains noticeable inconsistencies or incomplete realization of the specified setting.
        \item \textbf{2}: The scene has clear contradictions, unstable environmental cues, or weak alignment with the provided prompt.
        \item \textbf{1}: The scene is severely inconsistent, implausible, or confusing relative to the provided prompt.
    \end{itemize}

    \#\#\# 4. Important Constraints.\\
    \begin{itemize}[leftmargin=*, topsep=1pt, itemsep=0pt, parsep=0pt, partopsep=0pt]
        \item Evaluate \textbf{only} \textbf{Scene Consistency (SC)}.
        \item Use \textbf{Dialogue Context}, \textbf{Dialogue Scene.Location}, \textbf{Dialogue Scene.Time}, and \textbf{Dialogue Scene.Description} as the primary reference.
        \item Use \textbf{Dialogue Content} only as auxiliary evidence for whether the setting is appropriate.
        \item Do \textbf{not} score emotion, acting quality, character identity stability, turn-taking, or cinematography unless they directly affect scene coherence.
    \end{itemize}

    \#\#\# 5. Output Requirements.\\
    \begin{itemize}[leftmargin=*, topsep=1pt, itemsep=0pt, parsep=0pt, partopsep=0pt]
        \item Return \textbf{only} a single JSON object.
        \item The JSON must contain exactly two keys: \texttt{"score"} and \texttt{"rationale"}.
        \item \texttt{"score"} must be an integer from 1 to 5.
        \item \texttt{"rationale"} must explain the main evidence about scene coherence and alignment.
    \end{itemize}

    ---\textbf{TASK:} Evaluate the \texttt{Video} for \texttt{Scene Consistency (SC)} using the provided \texttt{Dialogue Context}, \texttt{Dialogue Scene}, and \texttt{Dialogue Content}, and return the JSON now.
    \end{systemprompt}
    \caption{The system prompt for evaluating scene consistency in generated audio-videos.}
    \label{fig:prompt_design_scene_consistency}
\end{figure*}

\begin{figure*}[t]
    \centering
    \setlength{\textfloatsep}{6pt}
    \begin{systemprompt}
    \small
    (Audio-Visual Evaluation Prompt: Character Consistency)\\
    \textit{You are a professional character continuity evaluator}. Your task is to evaluate whether the characters in an AI-generated 8--15 second audiovisual video remain visually stable, identifiable, and consistent with the provided generation prompt. Focus only on \textbf{Character Consistency (CC)}.\\

    \#\#\# 1. Input Variable Handling.\\
    You will receive data in the following format:
    \begin{itemize}[leftmargin=*, topsep=1pt, itemsep=0pt, parsep=0pt, partopsep=0pt]
        \item \textbf{Dialogue Context}: [Overall emotional state, mood, or situational context].
        \item \textbf{Dialogue Scene}: [Scene information, provided only as background].
        \item \textbf{Characters}: For each character, the following attributes may be provided:
        \begin{itemize}[leftmargin=1.5em, topsep=1pt, itemsep=0pt, parsep=0pt, partopsep=0pt]
            \item \textbf{Gender}
            \item \textbf{Age}
            \item \textbf{Ethnicity}
            \item \textbf{Appearance}
            \item \textbf{Emotion}
            \item \textbf{Action}
            \item \textbf{Outfit}
        \end{itemize}
        \item \textbf{Dialogue Content}: [The dialogue spoken in the video].
        \item \textbf{Video}: [An 8--15 second generated audio-video].
    \end{itemize}

    \#\#\# 2. Evaluation Focus.\\
    Evaluate whether each character in the \textbf{Video} remains visually and physically consistent with the \textbf{Characters} specification:\\
    \begin{itemize}[leftmargin=*, topsep=1pt, itemsep=0pt, parsep=0pt, partopsep=0pt]
        \item \textbf{Identity Stability}: Does each character maintain a stable face, body, and overall identity across the clip?
        \item \textbf{Attribute Consistency}: Are \textbf{Gender}, \textbf{Age}, \textbf{Ethnicity}, \textbf{Appearance}, and \textbf{Outfit} visually consistent with the prompt and stable over time?
        \item \textbf{Motion Robustness}: During speech, movement, or camera changes, does the same character remain recognizable without morphing or drift?
        \item \textbf{Cross-Moment Continuity}: If framing changes, do the characters remain the same identifiable individuals?
        \item \textbf{Absence of Character Artifacts}: Avoid face drift, clothing mutation, body inconsistency, or identity swapping.
    \end{itemize}

    \#\#\# 3. Scoring Standard (1--5).\\
    Assign an integer score from 1 to 5:
    \begin{itemize}[leftmargin=*, topsep=1pt, itemsep=0pt, parsep=0pt, partopsep=0pt]
        \item \textbf{5}: Character identity and appearance are fully stable and consistent with the \textbf{Characters} specification throughout the video.
        \item \textbf{4}: Characters are mostly consistent, with only minor continuity issues that do not affect recognition.
        \item \textbf{3}: Characters are recognizable overall, but noticeable instability or attribute drift appears in some moments.
        \item \textbf{2}: Character inconsistency is frequent, with visible identity drift or contradictory appearance cues.
        \item \textbf{1}: Characters are highly unstable or unreliable, with severe morphing, identity confusion, or broken continuity.
    \end{itemize}

    \#\#\# 4. Important Constraints.\\
    \begin{itemize}[leftmargin=*, topsep=1pt, itemsep=0pt, parsep=0pt, partopsep=0pt]
        \item Evaluate \textbf{only} \textbf{Character Consistency (CC)}.
        \item Use \textbf{Characters.Gender}, \textbf{Characters.Age}, \textbf{Characters.Ethnicity}, \textbf{Characters.Appearance}, and \textbf{Characters.Outfit} as the primary reference.
        \item Use \textbf{Characters.Action} and \textbf{Characters.Emotion} only as auxiliary cues if needed to interpret the character.
        \item Do \textbf{not} score emotional appropriateness, lip-sync, speaker assignment, turn-taking, or camera framing unless they directly affect character continuity.
    \end{itemize}

    \#\#\# 5. Output Requirements.\\
    \begin{itemize}[leftmargin=*, topsep=1pt, itemsep=0pt, parsep=0pt, partopsep=0pt]
        \item Return \textbf{only} a single JSON object.
        \item The JSON must contain exactly two keys: \texttt{"score"} and \texttt{"rationale"}.
        \item \texttt{"score"} must be an integer from 1 to 5.
        \item \texttt{"rationale"} must explain whether the characters remain visually stable and faithful to the provided \textbf{Characters} specification.
    \end{itemize}

    ---\textbf{TASK:} Evaluate the \texttt{Video} for \texttt{Character Consistency (CC)} using the provided \texttt{Characters} specification, and return the JSON now.
    \end{systemprompt}
    \caption{The system prompt for evaluating character consistency in generated audio-videos.}
    \label{fig:prompt_eval_character_consistency}
\end{figure*}

\begin{figure*}[t]
    \centering
    \setlength{\textfloatsep}{6pt}
    \begin{systemprompt}
    \small
    (Audio-Visual Evaluation Prompt: Speaker-Utterance Alignment)\\
    \textit{You are a professional dialogue alignment evaluator}. Your task is to evaluate whether the spoken utterances in an AI-generated 8--15 second audiovisual video are correctly aligned with the intended speaking characters. Focus only on \textbf{Speaker-Utterance Alignment (SA)}.\\

    \#\#\# 1. Input Variable Handling.\\
    You will receive data in the following format:
    \begin{itemize}[leftmargin=*, topsep=1pt, itemsep=0pt, parsep=0pt, partopsep=0pt]
        \item \textbf{Dialogue Context}: [Overall emotional state, mood, or situational context].
        \item \textbf{Dialogue Scene}: [Scene information, provided only as background].
        \item \textbf{Characters}: For each character, the following attributes may be provided:
        \begin{itemize}[leftmargin=1.5em, topsep=1pt, itemsep=0pt, parsep=0pt, partopsep=0pt]
            \item \textbf{Gender}
            \item \textbf{Age}
            \item \textbf{Ethnicity}
            \item \textbf{Appearance}
            \item \textbf{Emotion}
            \item \textbf{Action}
            \item \textbf{Outfit}
        \end{itemize}
        \item \textbf{Dialogue Content}: [The dialogue spoken in the video].
        \item \textbf{Video}: [An 8--15 second generated audio-video].
    \end{itemize}

    \#\#\# 2. Evaluation Focus.\\
    Evaluate whether the \textbf{Dialogue Content} is spoken by the correct visible character(s) in the \textbf{Video}:\\
    \begin{itemize}[leftmargin=*, topsep=1pt, itemsep=0pt, parsep=0pt, partopsep=0pt]
        \item \textbf{Speaker Attribution}: Is each utterance visually attributable to the correct character?
        \item \textbf{Dialogue-to-Character Matching}: Do the lines in \textbf{Dialogue Content} appear to be assigned to the intended speaker rather than swapped or ambiguously attached?
        \item \textbf{Visible Speaker Correspondence}: When a voice is heard, is there a plausible visible speaker in the \textbf{Video}?
        \item \textbf{Voice-Identity Consistency}: Does the same character appear to maintain a consistent speaking identity across the clip?
        \item \textbf{Multi-Speaker Clarity}: In multi-talker situations, is it clear who is speaking each line?
    \end{itemize}

    \#\#\# 3. Scoring Standard (1--5).\\
    Assign an integer score from 1 to 5:
    \begin{itemize}[leftmargin=*, topsep=1pt, itemsep=0pt, parsep=0pt, partopsep=0pt]
        \item \textbf{5}: All spoken content is clearly and correctly aligned with the intended visible speaker(s).
        \item \textbf{4}: Alignment is strong overall, with only minor ambiguity in a small moment.
        \item \textbf{3}: The main speaker assignment is understandable, but noticeable alignment issues or ambiguity remain.
        \item \textbf{2}: Multiple utterances appear weakly grounded, misassigned, or visually mismatched.
        \item \textbf{1}: Speaker-utterance correspondence is largely broken, making it difficult to tell who is speaking.
    \end{itemize}

    \#\#\# 4. Important Constraints.\\
    \begin{itemize}[leftmargin=*, topsep=1pt, itemsep=0pt, parsep=0pt, partopsep=0pt]
        \item Evaluate \textbf{only} \textbf{Speaker-Utterance Alignment (SA)}.
        \item Use \textbf{Dialogue Content} and \textbf{Characters} as the primary reference.
        \item You may use \textbf{Characters.Appearance}, \textbf{Characters.Action}, and visible speaking behavior to infer speaker identity.
        \item Do \textbf{not} score phoneme-level lip-sync precision unless it directly affects who appears to be speaking.
        \item Do \textbf{not} score emotional quality, scene realism, turn-taking structure, or cinematography.
    \end{itemize}

    \#\#\# 5. Output Requirements.\\
    \begin{itemize}[leftmargin=*, topsep=1pt, itemsep=0pt, parsep=0pt, partopsep=0pt]
        \item Return \textbf{only} a single JSON object.
        \item The JSON must contain exactly two keys: \texttt{"score"} and \texttt{"rationale"}.
        \item \texttt{"score"} must be an integer from 1 to 5.
        \item \texttt{"rationale"} must explain whether the spoken lines are aligned with the correct visible character(s).
    \end{itemize}

    ---\textbf{TASK:} Evaluate the \texttt{Video} for \texttt{Speaker-Utterance Alignment (SA)} using the provided \texttt{Dialogue Content} and \texttt{Characters}, and return the JSON now.
    \end{systemprompt}
    \caption{The system prompt for evaluating speaker-utterance alignment in generated audio-videos.}
    \label{fig:prompt_design_speaker_utterance_alignment}
\end{figure*}

\begin{figure*}[t]
    \centering
    \setlength{\textfloatsep}{6pt}
    \begin{systemprompt}
    \small
    (Audio-Visual Evaluation Prompt: Turn-Taking Logic)\\
    \textit{You are a professional conversational dynamics evaluator}. Your task is to evaluate whether an AI-generated 8--15 second audiovisual video presents natural and logically organized turn-taking between speakers. Focus only on \textbf{Turn-Taking Logic (TT)}.\\

    \#\#\# 1. Input Variable Handling.\\
    You will receive data in the following format:
    \begin{itemize}[leftmargin=*, topsep=1pt, itemsep=0pt, parsep=0pt, partopsep=0pt]
        \item \textbf{Dialogue Context}: [Overall emotional state, mood, or situational context].
        \item \textbf{Dialogue Scene}: [Scene information, provided only as background].
        \item \textbf{Characters}: [Character specifications, provided for identifying speaker transitions when needed].
        \item \textbf{Dialogue Content}: [The dialogue spoken in the video].
        \item \textbf{Video}: [An 8--15 second generated audio-video].
    \end{itemize}

    \#\#\# 2. Evaluation Focus.\\
    Evaluate whether the \textbf{Dialogue Content} unfolds in the \textbf{Video} with natural conversational timing and logical turn organization:\\
    \begin{itemize}[leftmargin=*, topsep=1pt, itemsep=0pt, parsep=0pt, partopsep=0pt]
        \item \textbf{Turn Order}: Do the utterances occur in a sensible order consistent with the provided \textbf{Dialogue Content}?
        \item \textbf{Turn Boundary Clarity}: Are speaker transitions clear rather than merged, truncated, or confusing?
        \item \textbf{Interaction Rhythm}: Do pauses, reactions, interruptions, and response gaps feel natural for a real conversation?
        \item \textbf{Response Coherence}: Does each turn seem to follow the previous one logically?
        \item \textbf{Structural Integrity}: Avoid missing turns, repeated turns, unnatural overlap, delayed responses, or abrupt discontinuities in the dialogue flow.
    \end{itemize}

    \#\#\# 3. Scoring Standard (1--5).\\
    Assign an integer score from 1 to 5:
    \begin{itemize}[leftmargin=*, topsep=1pt, itemsep=0pt, parsep=0pt, partopsep=0pt]
        \item \textbf{5}: Turn-taking is fully natural, temporally well-structured, and easy to follow.
        \item \textbf{4}: Turn-taking is mostly natural, with only slight timing or transition issues.
        \item \textbf{3}: The interaction structure is understandable, but noticeable pacing or transition problems remain.
        \item \textbf{2}: Turn-taking is frequently awkward, confusing, truncated, or poorly organized.
        \item \textbf{1}: Conversational structure is severely broken, making the interaction hard to follow.
    \end{itemize}

    \#\#\# 4. Important Constraints.\\
    \begin{itemize}[leftmargin=*, topsep=1pt, itemsep=0pt, parsep=0pt, partopsep=0pt]
        \item Evaluate \textbf{only} \textbf{Turn-Taking Logic (TT)}.
        \item Use \textbf{Dialogue Content} as the primary reference.
        \item Use \textbf{Characters} only when needed to determine who is taking each turn.
        \item Do \textbf{not} score lip-sync quality, character identity stability, emotional appropriateness, or camera framing unless they directly affect turn organization.
        \item Focus on temporal and conversational structure rather than semantic richness.
    \end{itemize}

    \#\#\# 5. Output Requirements.\\
    \begin{itemize}[leftmargin=*, topsep=1pt, itemsep=0pt, parsep=0pt, partopsep=0pt]
        \item Return \textbf{only} a single JSON object.
        \item The JSON must contain exactly two keys: \texttt{"score"} and \texttt{"rationale"}.
        \item \texttt{"score"} must be an integer from 1 to 5.
        \item \texttt{"rationale"} must explain whether the turn-taking structure is natural and logically organized.
    \end{itemize}

    ---\textbf{TASK:} Evaluate the \texttt{Video} for \texttt{Turn-Taking Logic (TT)} using the provided \texttt{Dialogue Content}, and return the JSON now.
    \end{systemprompt}
    \caption{The system prompt for evaluating turn-taking logic in generated audio-videos.}
    \label{fig:prompt_design_turn_taking_logic}
\end{figure*}

\begin{figure*}[t]
    \centering
    \setlength{\textfloatsep}{6pt}
    \begin{systemprompt}
    \small
    (Audio-Visual Evaluation Prompt: Affective \& Expressive Alignment)\\
    \textit{You are a professional affective performance evaluator}. Your task is to evaluate whether the emotional and expressive performance in an AI-generated 8--15 second audiovisual video aligns with the provided generation prompt. Focus only on \textbf{Affective \& Expressive Alignment (EA)}.\\

    \#\#\# 1. Input Variable Handling.\\
    You will receive data in the following format:
    \begin{itemize}[leftmargin=*, topsep=1pt, itemsep=0pt, parsep=0pt, partopsep=0pt]
        \item \textbf{Dialogue Context}: [Overall emotional state, mood, or situational context].
        \item \textbf{Dialogue Scene}: [Scene information, provided only as background].
        \item \textbf{Characters}: For each character, the following attributes may be provided:
        \begin{itemize}[leftmargin=1.5em, topsep=1pt, itemsep=0pt, parsep=0pt, partopsep=0pt]
            \item \textbf{Gender}
            \item \textbf{Age}
            \item \textbf{Ethnicity}
            \item \textbf{Appearance}
            \item \textbf{Emotion}
            \item \textbf{Action}
            \item \textbf{Outfit}
        \end{itemize}
        \item \textbf{Dialogue Content}: [The dialogue spoken in the video].
        \item \textbf{Video}: [An 8--15 second generated audio-video].
    \end{itemize}

    \#\#\# 2. Evaluation Focus.\\
    Evaluate whether the emotional delivery and visible expressive behavior in the \textbf{Video} match the intended affective cues in \textbf{Dialogue Context}, \textbf{Characters}, and \textbf{Dialogue Content}:\\
    \begin{itemize}[leftmargin=*, topsep=1pt, itemsep=0pt, parsep=0pt, partopsep=0pt]
        \item \textbf{Context-Level Emotion Match}: Does the overall emotional tone match \textbf{Dialogue Context}?
        \item \textbf{Character-Level Expression Match}: Do facial expressions, body language, vocal tone, and delivery style match \textbf{Characters.Emotion} and \textbf{Characters.Action}?
        \item \textbf{Dialogue-Level Expressiveness}: Does the way each line is delivered fit the meaning of the \textbf{Dialogue Content}?
        \item \textbf{Cross-Modal Affective Coherence}: Do voice, facial expression, and body movement reinforce the same feeling?
        \item \textbf{No Affective Contradiction}: Avoid mismatches such as cheerful delivery in a sad exchange, flat affect in an intense confrontation, or body language that contradicts the spoken line.
    \end{itemize}

    \#\#\# 3. Scoring Standard (1--5).\\
    Assign an integer score from 1 to 5:
    \begin{itemize}[leftmargin=*, topsep=1pt, itemsep=0pt, parsep=0pt, partopsep=0pt]
        \item \textbf{5}: Emotional and expressive behavior is highly natural, precise, and fully aligned with the provided prompt.
        \item \textbf{4}: Expression is strong and appropriate overall, with only minor mismatch or stiffness.
        \item \textbf{3}: Emotional intent is partially conveyed, but the performance is generic, uneven, or only moderately aligned.
        \item \textbf{2}: Expression is noticeably weak, mismatched, or confusing relative to the prompt.
        \item \textbf{1}: Emotional and expressive behavior strongly contradicts the intended affect or meaning.
    \end{itemize}

    \#\#\# 4. Important Constraints.\\
    \begin{itemize}[leftmargin=*, topsep=1pt, itemsep=0pt, parsep=0pt, partopsep=0pt]
        \item Evaluate \textbf{only} \textbf{Affective \& Expressive Alignment (EA)}.
        \item Use \textbf{Dialogue Context}, \textbf{Characters.Emotion}, \textbf{Characters.Action}, and \textbf{Dialogue Content} as the primary reference.
        \item Do \textbf{not} score scene plausibility, character identity continuity, turn-taking organization, or camera framing unless they directly interfere with emotional expression.
        \item Focus on whether the audiovisual performance conveys the intended feeling.
    \end{itemize}

    \#\#\# 5. Output Requirements.\\
    \begin{itemize}[leftmargin=*, topsep=1pt, itemsep=0pt, parsep=0pt, partopsep=0pt]
        \item Return \textbf{only} a single JSON object.
        \item The JSON must contain exactly two keys: \texttt{"score"} and \texttt{"rationale"}.
        \item \texttt{"score"} must be an integer from 1 to 5.
        \item \texttt{"rationale"} must explain how well the emotional and expressive behavior aligns with the provided prompt.
    \end{itemize}

    ---\textbf{TASK:} Evaluate the \texttt{Video} for \texttt{Affective \& Expressive Alignment (EA)} using the provided \texttt{Dialogue Context}, \texttt{Characters}, and \texttt{Dialogue Content}, and return the JSON now.
    \end{systemprompt}
    \caption{The system prompt for evaluating affective and expressive alignment in generated audio-videos.}
    \label{fig:prompt_design_affective_expressive_alignment}
\end{figure*}

\begin{figure*}[t]
    \centering
    \setlength{\textfloatsep}{6pt}
    \begin{systemprompt}
    \small
    (Audio-Visual Evaluation Prompt: Camera / Composition Alignment)\\
    \textit{You are a professional cinematography evaluator}. Your task is to evaluate whether an AI-generated 8--15 second audiovisual video uses framing, composition, and camera attention in a way that appropriately supports the dialogue and the active speaker. Focus only on \textbf{Camera / Composition Alignment (CA)}.\\

    \#\#\# 1. Input Variable Handling.\\
    You will receive data in the following format:
    \begin{itemize}[leftmargin=*, topsep=1pt, itemsep=0pt, parsep=0pt, partopsep=0pt]
        \item \textbf{Dialogue Context}: [Overall emotional state, mood, or situational context].
        \item \textbf{Dialogue Scene}:
        \begin{itemize}[leftmargin=1.5em, topsep=1pt, itemsep=0pt, parsep=0pt, partopsep=0pt]
            \item \textbf{Location}
            \item \textbf{Time}
            \item \textbf{Description}
        \end{itemize}
        \item \textbf{Characters}: For each character, the following attributes may be provided:
        \begin{itemize}[leftmargin=1.5em, topsep=1pt, itemsep=0pt, parsep=0pt, partopsep=0pt]
            \item \textbf{Gender}
            \item \textbf{Age}
            \item \textbf{Ethnicity}
            \item \textbf{Appearance}
            \item \textbf{Emotion}
            \item \textbf{Action}
            \item \textbf{Outfit}
        \end{itemize}
        \item \textbf{Dialogue Content}: [The dialogue spoken in the video].
        \item \textbf{Video}: [An 8--15 second generated audio-video].
    \end{itemize}

    \#\#\# 2. Evaluation Focus.\\
    Evaluate whether framing and composition in the \textbf{Video} appropriately support the active moment in the dialogue:\\
    \begin{itemize}[leftmargin=*, topsep=1pt, itemsep=0pt, parsep=0pt, partopsep=0pt]
        \item \textbf{Speaker-Centric Focus}: When a character is speaking, is that character visually emphasized or given appropriate prominence?
        \item \textbf{Dialogue-Aware Framing}: Does the camera respond appropriately when the speaking turn or interaction focus changes?
        \item \textbf{Composition Clarity}: Do subject placement, shot scale, and framing help the viewer understand who is important at each moment?
        \item \textbf{Action Visibility}: Are key facial expressions, gestures, or interaction cues visible when needed, especially those implied by \textbf{Characters.Action} and \textbf{Dialogue Content}?
        \item \textbf{No Distracting Misframing}: Avoid centering the wrong person, hiding the speaker, awkward cropping, or compositions that reduce conversational readability.
    \end{itemize}

    \#\#\# 3. Scoring Standard (1--5).\\
    Assign an integer score from 1 to 5:
    \begin{itemize}[leftmargin=*, topsep=1pt, itemsep=0pt, parsep=0pt, partopsep=0pt]
        \item \textbf{5}: Camera framing and composition consistently and effectively support the active speaker and dialogue moment.
        \item \textbf{4}: Framing is strong overall, with only small issues that do not substantially harm clarity.
        \item \textbf{3}: Viewer attention is generally understandable, but composition is sometimes generic, delayed, or only partially supportive.
        \item \textbf{2}: Framing frequently misdirects attention or fails to emphasize the relevant subject.
        \item \textbf{1}: Camera/composition alignment is severely broken, making it difficult to follow who or what should be visually prioritized.
    \end{itemize}

    \#\#\# 4. Important Constraints.\\
    \begin{itemize}[leftmargin=*, topsep=1pt, itemsep=0pt, parsep=0pt, partopsep=0pt]
        \item Evaluate \textbf{only} \textbf{Camera / Composition Alignment (CA)}.
        \item Use \textbf{Dialogue Content} as the primary cue for who should be visually prioritized.
        \item Use \textbf{Characters.Action}, visible speaker behavior, and interaction cues as secondary evidence.
        \item Use \textbf{Dialogue Scene} only when needed to judge whether composition supports the scene setup.
        \item Do \textbf{not} score scene realism, emotional appropriateness, character identity continuity, or turn-taking except when they directly affect framing relevance.
    \end{itemize}

    \#\#\# 5. Output Requirements.\\
    \begin{itemize}[leftmargin=*, topsep=1pt, itemsep=0pt, parsep=0pt, partopsep=0pt]
        \item Return \textbf{only} a single JSON object.
        \item The JSON must contain exactly two keys: \texttt{"score"} and \texttt{"rationale"}.
        \item \texttt{"score"} must be an integer from 1 to 5.
        \item \texttt{"rationale"} must explain whether the framing and composition properly support the active speaker and dialogue moment.
    \end{itemize}

    ---\textbf{TASK:} Evaluate the \texttt{Video} for \texttt{Camera / Composition Alignment (CA)} using the provided \texttt{Dialogue Content}, \texttt{Characters}, and \texttt{Dialogue Scene}, and return the JSON now.
    \end{systemprompt}
    \caption{The system prompt for evaluating camera and composition alignment in generated audio-videos.}
    \label{fig:prompt_design_camera_composition_alignment}
\end{figure*}

\end{document}